\documentclass[superscriptaddress,groupedaddress,nofootnoteinbib,11pt]{article}
\pdfoutput=1 
\usepackage{jheppub}

\usepackage[utf8x]{inputenc}
\usepackage{mathtools,slashed,mathrsfs}
\usepackage[caption=false]{subfig}
\usepackage{dcolumn}
\usepackage{multirow}
\usepackage{tabularx}
\usepackage{float}
\usepackage{booktabs}
\usepackage{bm}
\usepackage{comment}
\usepackage{setspace}
\usepackage[dvipsnames]{xcolor}
\usepackage[normalem]{ulem} 
\usepackage{enumerate}
\usepackage{siunitx}
\graphicspath{{../Mathematica/}}

\def\DD{\mathcal{D}}
\def\ZZ{\mathbf{Z}}
\newcommand\identity{1\kern-0.25em\text{l}}

\def\LL{\mathcal{L}}
\def\mat{\begin{pmatrix}}
\def\rix{\end{pmatrix}}
\def\ket{\right \rangle}
\def\bra{\left\langle}
\def\rar{\rightarrow}

\def\dag{\dagger}

\def\({\left(}
\def\){\right)}
\def\[{\left[}
\def\]{\right]}

\def\where{\quad \text{where}\quad}

\def\OO{\mathcal{O}}

\newcommand\nn{\nonumber}
\newcommand\bea{\begin{equation}}
\newcommand\eea{\end{equation}}
\def\l{\left(}
\def\r{\right)}

\begin{document}

\title{Dyonic bound states}

\date{\today}
\author{Anson Hook,}
\author{Clayton Ristow}

\affiliation{Maryland Center for Fundamental Physics, University of Maryland, College Park, MD 20742, U.S.A.}

\emailAdd{hook@umd.edu}
\emailAdd{cristow@umd.edu}

\abstract{
We study (multi) fermion - monopole bound states, many of which are the states that dyons adiabatically transition into as fermions become light.  The properties of these bound states depend critically on the UV symmetries preserved by the fermion mass terms, their relative size, and the value of $\theta$.  Depending on the relative size of the mass terms and the value of $\theta$, the bound states can undergo phase transitions as well as transition from being stable to unstable.  In some simple situations, the bound state solution can be related to the Witten effect of another theory with fewer fermions and larger gauge coupling.  These bound states are a result of mass terms and symmetry breaking boundary conditions at the monopole core and, consequently, these bound states do not necessarily have definite quantum numbers under accidental IR symmetries.  Additionally, they have binding energies that are $\mathcal{O}(1)$ times the fermion mass and bound state radii of order their inverse mass.
As the massless limit is approached, the bound state radii approach infinity, and they become new asymptotic states with odd quantum numbers giving a dynamical understanding to the origin of semitons.  
}

\maketitle

\section{Introduction}

Monopoles have been an intriguing idea ever since the discovery of Maxwell's equations.  Their existence would render Maxwell's equations much more symmetric, as well as explain interesting observed phenomena such as the quantization of electric charge.  Monopoles really came to the forefront when it was found that they were a consequence of many spontaneously broken gauge theories in the form of the 't Hooft-Polyakov monopole~\cite{tHooft:1974kcl,Polyakov:1974ek}.

Periodicity of the $\theta F \tilde F$ coupling requires that monopoles come with an infinite number of additional states with both electric and magnetic charge, the dyons.  Sending $\theta \rightarrow \theta + 2 \pi$ exchanges monopoles with dyons and dyons with dyons of one unit of electric charge less.  
Originally, these dyonic excitations were found by quantizing the charge rotator degree of freedom of the 't Hooft-Polyakov monopole, but dyons are often colloquially interpreted as charged $W$ bosons bound to the monopole.

The story changes once light fermions are added to the mix as fermions, via the chiral anomaly, can carry topological charge.  It becomes energetically favorable to excite the light charged fermions as opposed to the charge rotator/W boson~\cite{Callan:1982ah,Preskill:1984gd}.  In particular, as one adiabatically decreases fermion masses, the many dyons smoothly become either multi-fermion bound states or monopoles with asymptotic fermions.  We will show that these multi-fermion bound states have binding energy of order the mass of the fermion and radii of order inverse the fermion mass.  These bound states are in contrast to previously studied monopole - single fermion bound states utilizing non-trivial electric and magnetic moments to exist~\cite{Olaussen:1983qc,Osland:1984yu,Osland:1984ys,Osland:1985va}, hydrogen-like bound states stabilized by electromagnetism~\cite{Zhang:1988ab} or the Higgs mediated force~\cite{Tang:1982fc}, and dyon-monopole bound states in supersymmetric theories~\cite{Sen:1994yi,Sethi:1995zm,Gauntlett:1995fu}.  Our bound states are most similar to the zero-mode bound states required by index theorems, which can be interpreted as bound states with $E = m$~\cite{Jackiw:1975fn,Callias:1977kg,Weinberg:1979ma}.  Unlike many of the previous examples, there is no long range force supporting our bound states and it is supported entirely by the Callan-Rubakov symmetry breaking boundary conditions.  Our bound states are not required by index theorems and only have $ E \approx m$.  Because they are supported by boundary conditions that break the IR symmetries, but preserve the UV symmetries, these bound states do not have well defined quantum numbers under the approximate IR symmetries but have well defined quantum numbers under the exact UV symmetries~\footnote{A heuristic understanding of the 2 fermion bound state is to take the 2-to-2 fermion flavor breaking Callan-Rubakov scattering process ( say $u + d$ to $\overline e + \overline u$) and work in the limit that the 2 fermions do not have enough energy to escape to asymptotic infinity.  The scattering process repeats itself periodically converting the fermions back and forth into each other, resulting in a bound state with ill defined quantum numbers under the IR symmetries.}.

In this article, we conduct a systematic study of these (multi)-fermion bound states.  
The IR theory we consider consists of $N_f$ Dirac fermions with equal masses and charges and an $SU(N_f)$ flavor symmetry.  In the UV, this $U(1)$ gauge theory is embedded into a spontaneously broken $SU(2)$ gauge theory.
From the UV perspective, there are two types of mass terms, those that preserve either an $SO(N_f)$ or $Sp(N_f)$ flavor symmetry, each with their own unique phenomenology.  For example, in the $SO(N_f)$ case, $\theta$ is only $\pi$ periodic, as opposed to usual $2 \pi$ periodicity.  In each of these scenarios, we numerically and analytically study the charge distribution of the bound states and their associated energetics.  There are many different bound states whose stability varies with $\theta$.  For example, when $\theta = \pi$ in an $Sp(N_f)$ flavor symmetric theory, the lowest energy magnetically charged state is not the $\theta$-charged monopole, but instead a fermion - monopole bound state.
When the masses become unequal, the dynamics change.  As an example, bound states can become unstable to decay.

In an $SO$ ($Sp$) symmetric theory, there exist multi-fermion bound states when $\theta=0$ until one has bound $\approx 0.6 N_f$ ($\approx 0.2 N_f$) fermions onto the monopole.  Near the critical point, the bound states become unstable via a process similar to nuclear decay, as the unstable states decay via fermions tunneling away from bound states that are classically stable but quantum mechanically unstable.  In the $Sp$ symmetric theory, multi-fermion flavorless bound states only exist when there is an even number of fermions, while flavorful bound states only exist as a possibility as single-fermion bound states.  Meanwhile, in $SO$ symmetric theories, the multi-fermion flavorless bound states exist for an even or odd number of fermions.


One cannot in good conscience discuss all of these effects without commenting on GUT monopoles.  We discuss how the $\theta$ dependence of $SU(5)$ monopoles does not depend on the lightest charged particle, the electron, and instead depends on the mass of heavier particles.  
Unfortunately, the possible existence of bound states cannot be said for certain due to confinement.  Finally, we give an estimate for the velocity at which the Callan-Rubakov effect becomes unsuppressed.  Namely, we estimate that a proton needs to be shot with a speed of $v \sim 2 \times 10^{-2}$ at an $SU(5)$ GUT monopole in order for its conversion to a positron to become unsuppressed and saturate unitarity.

We conclude with a few comments on how our results help provide a better understanding of the massless fermion limit.  As the mass of the fermions is taken to zero, the bound states become larger and larger and eventually become new asymptotic states.  Since the bound states are stabilized by a symmetry breaking boundary condition, these bound states, and the asymptotic states that they become, do not carry good IR symmetries.  These new, slightly odd asymptotic states are called semitons~\cite{Callan:1983tm} and our approach gives physical understanding to how they appear in the massless limit.

In Sec.~\ref{Sec: groundwork}, we describe the UV theories of interest and their bosonized form that we eventually study numerically.  The details of bosonization are relegated to Apps.~\ref{Sec: Bosonization Details} and~\ref{App: boson}.
We ease into the discussion of bound states by studying an $N_f = 1$ example in Sec.~\ref{Sec: Nf1}.
In Sec.~\ref{Sec: Nf2} we discuss bound states in the context of $N_f = 2$ and discuss the $N_f \ge 4$ case in Sec.~\ref{Sec: Nfgeq4}.
For fun, in Sec.~\ref{Sec: SU5} we discuss $SU(5)$ monopoles.  We discuss the relationship of our bound states to the semiton puzzle in Sec.~\ref{Sec: semiton}.  Finally, we conclude in Sec.~\ref{Sec: conclusion}.

\section{UV Theories and Their Bosonization} \label{Sec: groundwork}

In this section, we describe the theories we study and their bosonized form.  We are studying an $SU(2)$ gauge theory with an adjoint scalar $\Phi$ and $N_f$ Weyl fermions, $\psi_i$ with $i = 1 \cdots N_f$, in the fundamental representation ($N_f$ is required to be even by the Witten anomaly~\cite{Witten:1982fp}).  The scalar $\Phi$ obtains a vev $v$ and breaks the theory down to a $U(1)$ giving a 't Hooft-Polyakov monopole.

\paragraph{$SO(N_f)$ Flavor Symmetries}: 
There are two different mass terms for our fermions that each give different results.  The first is a mass term of the form
\bea \label{Eq: somassterm}
\delta \mathcal{L} = - \frac{y}{2} \psi_i^a \Phi_c^b \psi_i^c \epsilon_{ab},
\eea
where $i$ is the flavor index and $a,b,c$ are the gauge indices.  
Mass terms of this form break the $SU(N_f)$ flavor symmetry down to an $SO(N_f)$ flavor symmetry.  Eventually we will allow the masses to be unequal, thus breaking the $SO(N_f)$ flavor symmetry, but even in that case, we will still call it the $SO(N_f)$ theory to indicate that the mass terms are of the form shown in Eq.~\ref{Eq: somassterm}. 

To simplify our numerical analysis, we will follow in the footsteps of Ref.~\cite{Callan:1982ah,Callan:1982au} with details given in Apps.~\ref{Sec: Bosonization Details} and~\ref{App: boson}.  Firstly, we first restrict ourselves to the $J=0$ state as all higher momentum states have higher energy due to non-zero angular momentum.  This process reduces our 4-dimensional starting point to physics on a half line.  To study the remaining problem, we utilize the well-known fact that a 2D fermion is equivalent to a 2D boson.  Integrating out the photon, the end result is a theory with the Lagrangian
\bea \label{Eq: SON Lag}
4 \pi \mathcal{L} =  \frac{1}{2} \sum_i \l \partial \phi_i \r^2 + \l \frac{\pi m(r)}{2} \r^2 \sum_i \cos \l \phi_i \r - \frac{\alpha}{2 \pi r^2} \l \sum_i \frac{1}{2} \phi_i - \theta \r^2 ,
\eea
where $\alpha = g^2/4 \pi$ is the fine structure constant for the $SU(2)$ and $U(1)$ gauge theories.
The mass term has the limit $m^2(\infty) = m^2$, where $m = y v$ is the mass of the fermion and has the analytic form $m^2(r) = e^{K_0(\pi^2 e^{-\gamma} m r/4)} m^2$.  Additionally, all phases have been rotated into $\theta$.  The derivation of Eq.~\ref{Eq: SON Lag} involved integration by parts so that the Lagrangian is not explicitly invariant under $\theta \rightarrow \theta + 2 \pi$ or $\phi_i \rightarrow \phi_i + 4 \pi$.  Instead, when $\theta = 2 \pi$, the object under consideration is the dyon of the theory as opposed to the $\theta = 0$ monopole.

Bosonization maps solitons of $\phi_i$ to the fermion $\psi_i = \begin{pmatrix}
           \psi_i^1 \\
           \psi_i^2
         \end{pmatrix}$ of the original picture.  
While we will largely not be concerning ourselves with questions of dynamics, it is still useful to dictate the matching of solitons and fermions.  A soliton solution is $2 \pi$ at infinity and $0$ at the origin, while an anti-soliton solution is $0$ at infinity and $2 \pi$ at the origin.  An incoming $\phi_i$ soliton is an incoming $\psi_i^1$ left-handed particle, while an incoming $\phi_i$ anti-soliton is a $\psi_i^{1,\dagger}$ right-handed anti-particle.
An outgoing $\phi_i$ soliton is an outgoing $\psi_i^{2,\dagger}$  right-handed particle, while an outgoing $\phi_i$ anti-soliton is a $\psi_i^2$ left-handed anti-particle.  

The final ingredients are the boundary conditions which, when combined with the mapping described in the previous paragraph, determine the flavor symmetry of the problem.  The boundary conditions are
\bea
\label{Eq: SON boundary}
\partial_r \phi_i(r=0) = 0 .
\eea
We are now at a state where we can study any bound state in the theory by simply solving Eq.~\ref{Eq: SON Lag} subject to the boundary conditions shown in Eq.~\ref{Eq: SON boundary}.

One important point to note about this theory is that $\theta$ is $\pi$ periodic.  At the level of the 4D fermionic path integral, one can make a field redefinition $\psi_1 \rightarrow -\psi_1$.  Because of the Majorana-like mass term, the Lagrangian remains unchanged other than the anomaly, which sends $\theta \rightarrow \theta + \pi$.  Since a field redefinition cannot change the physics, the two path integrals are equivalent and the physics at $\theta$ and $\theta + \pi$ are equivalent.  At the 2D bosonized level, $\theta$ being $\pi$ periodic can be similarly seen by the transformation $\theta \rightarrow \theta - \pi$ combined with $\phi_i \rightarrow \phi_i  -2 \pi$.  The fact that this $SO$ theory does not have a $B-L$ symmetry and $\theta = \theta + \pi$ will be important for the dyonic spectrum of the theory.

\paragraph{$Sp(N_f)$ Flavor Symmetries} : 
The second mass term we consider is of the form
\bea
\delta \mathcal{L} = - m \psi_i^a \psi_j^b \epsilon_{ab} \epsilon^{ij}.
\eea
A mass term of this form breaks the $SU(N_f)$ flavor symmetry down to $Sp(N_f)$~\footnote{Due to the center of the $SU(2)$ gauge symmetry, the flavor symmetry is actually $PSp(N_f)$.}.  As before, when we allow for mass terms to be unequal using $\delta \mathcal{L} = y \psi \Phi \psi$, we will still refer to this theory as the $Sp(N_f)$ theory to indicate how its mass terms are written.

The process of bosonization keeps only an $SU(2)^{N_f/2}$ subgroup of $Sp(N_f)$ manifest.  We separate out $N_f/2$ pairs of fermions and bosonize each of these pairs together. 
Because we separated our theory into pairs of fermions and bosonized it, we get pairs of scalars, $\phi_{b,i}$ and $\phi_{\ell,i}$ where $i$ now only goes from 1 to $N_f/2$ and $b/\ell$ stands for bosonized bosons or leptons.  For simplicity of labeling, we will group the $i$-th fermion with the $N_f - i + 1$-th fermion.  The Lagrangian for these scalars is
\begin{eqnarray} \label{Eq: SpN Lag}
4 \pi \mathcal{L}&=&  \frac{1}{2} \sum_i \l \partial \phi_{b,i} \r^2 + \frac{1}{2} \sum_i \l \partial \phi_{\ell,i} \r^2 + \l \frac{\pi m}{2} \r^2 \sum_i \cos \phi_{b,i} + \l \frac{\pi m}{2} \r^2 \sum_i \cos \phi_{\ell,i} \nn \\
&-& \frac{\alpha}{2 \pi r^2} \l \sum_i \frac{1}{2} \phi_{b,i} + \sum_i \frac{1}{2} \phi_{\ell,i} - \theta \r^2 ,
\end{eqnarray}
where all phases have been rotated into $\theta$ as before.
Up to relabeling and moving far from the monopole ($r \gg 1/m$), this bosonized Lagrangian is identical to Eq.~\ref{Eq: SON Lag} as the IR Lagrangian cannot tell the difference between these two UV theories.

As before, we can map the solitons of this bosonized theory back to the original fermions.  An incoming $\phi_{b,i}$ ($\phi_{\ell,i}$) soliton is an incoming $\psi_i^1$ ($\psi_i^2$) left-handed particle, while an incoming $\phi_{b,i}$ ($\phi_{\ell,i}$) anti-soliton is a $\psi_{N_f - i + 1}^1$ ($\psi_{N_f - i + 1}^2$) left-handed anti-particle.  An outgoing $\phi_{b,i}$ ($\phi_{\ell,i}$) soliton is an outgoing $\psi_{N_f - i + 1}^{1,\dagger}$ ($\psi_{N_f - i + 1}^{2,\dagger}$)  right-handed particle, while an outgoing $\phi_{b,i}$ ($\phi_{\ell,i}$) anti-soliton is a $\psi_i^{1,\dagger}$ ($\psi_i^{2,\dagger}$) right-handed anti-particle.

Mirroring the previous case, the boundary conditions are what determine the UV flavor symmetries.  In this case, the boundary conditions are 
\bea \label{Eq: SpN boundary}
\phi_{b,i}(r=0) = \phi_{\ell,i}(r=0) \, \text{mod} \, 4 \pi \qquad \partial_r \phi_{b,i}(r=0) = - \partial_r \phi_{\ell,i}(r=0).
\eea
The first of these boundary conditions is what imposes quantization of the $U(1)$ charges of the $Sp(N_f)$ Cartan subalgebra.
Now, like the $SO(N_f)$ example, we can study any bound state in the theory by simply solving Eq.~\ref{Eq: SpN Lag} subject to the boundary conditions shown in Eq.~\ref{Eq: SpN boundary}.

\section{An $N_f =1$ example} \label{Sec: Nf1}

The examples that we will discuss in later sections will be rather complicated, so we begin with an $N_f = 1$ example.  Many of the $N_f = 1$ phenomena will generalize to larger $N_f$.  Additionally, we will find some analytic approximations for the bound states and their energies that will prove to be very useful when considering $N_f>1$.

\begin{figure}[t]
 \centering
 \includegraphics[width=0.6\textwidth]{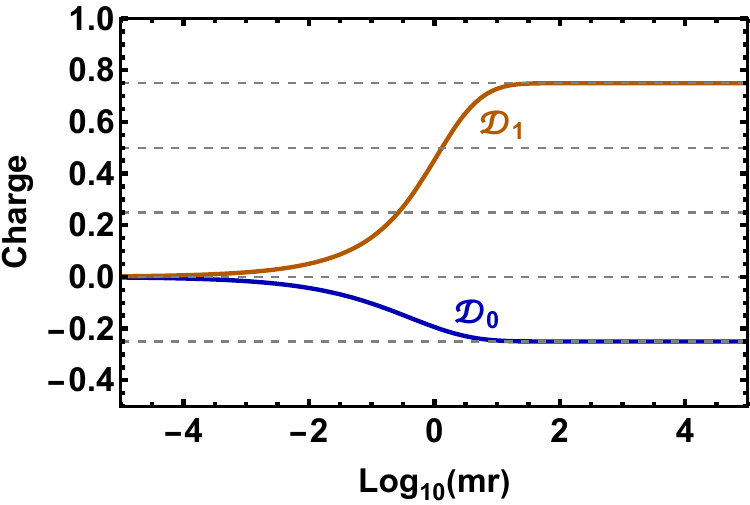}
 \caption{An example of the two potentially stable dyonic bound states in our $N_f=1$ toy model for $\alpha=0.1$ and $\theta=\pi/2$. These states are plotted in terms of $Q(r)=\frac{\phi(r)-\phi(0)}{2\pi}$, the total charge enclosed in a radius $r$ of the monopole core.}
 \label{Fig: Nf1 Bound States}
\end{figure}

There are several ways to arrive at our $N_f = 1$ theory from a more realistic theory. The simplest way is to start with an $N_f =2$ theory with $SO(2)$ symmetric mass terms and decouple one of the two fermions by increasing its mass. Another is to take any $N_f$ and look for a solution where all of the bosonized scalar profiles are identical, a solution that exists and is important for both $SO(N_f)$ and $Sp(N_f)$ flavor symmetries, albeit with slightly different $\alpha$ and $\theta$ in the $N_f=1$ theory than in the original theory. Regardless, the Lagrangian for the $N_f=1$ theory can be expressed as
\bea
\label{Eq: Bosonized Lagrangian}
L=\frac{1}{4\pi}\int_0^\infty dr\; \frac{1}{2}\partial_\mu\phi\partial^\mu\phi  - \(\frac{\pi m}{2}\)^2(1-\cos(\phi))-\frac{\alpha}{8\pi r^2}\(\phi-\theta\)^2
\eea
with boundary conditions $\partial_r\phi=0$ at $r=0$.
The time-independent equation of motion is 
\bea
\label{Eq: EoM Nf1}
\partial_r^2 \phi = \frac{\pi^2 m^2}{4} \sin \phi +\frac{\alpha}{4\pi r^2} ( \phi - \theta) .
\eea
The bound states we are after are defined to be solutions to this equation of motion of minimal energy.  Demanding that our solutions have finite energy requires $\phi(r=0)=\theta$ and $\phi(r=\infty)=2\pi q$, where $q$ is an integer.  That means there is an infinite tower of potential bound states, which we denote $\DD_q$, one for each choice of $q$, with total electric charge $Q_{EM}^{tot}=q-\theta/2\pi$.  It is easy to see that for states with $|Q_{EM}^{tot}|>1$, the energy will be minimized by putting some number of solitons at $r=\infty$, which can hardly be considered a bound state. Thus, in our $N_f=1$ theory, there are only two potentially stable bound-state dyons for $0<\theta<2\pi$: $\DD_0$ and $\DD_1$.  An example of these states is shown in Fig.~\ref{Fig: Nf1 Bound States} in terms of the total charge ($Q(r)=\frac{\phi(r)-\phi(0)}{2\pi}$) enclosed in a radius $r$.  The mapping of the global and gauge charges into the bosonic language can be found in App.~\ref{Appendix: currents}.

We can find useful analytic approximations to this solution in the $mr\ll1$ and $mr\gg 1$ limits. For $mr\ll1$, the mass term in Eq.~\ref{Eq: EoM Nf1} can be neglected and we naturally have the trivial solution $\phi=\theta$. However, as we will see, for theories with $N_f>1$, $\phi$ is occasionally forced to take values other than this trivial one in the $mr\ll1$ regime, so it will be important to consider other solutions. Say, for example, that $\phi$ must satisfy $\phi(r=r_0)=\phi_0$. Then we can solve Eq.~\ref{Eq: EoM Nf1} neglecting the mass term to find
\bea
\label{Eq: massless asymp}
\phi \approx \theta+(\phi_0-\theta)\(\frac{r}{r_0}\)^{-\beta}\where \beta=\frac{1}{2}\(\sqrt{1+\frac{\alpha}{\pi}}-1\)\approx\frac{\alpha}{4\pi} .
\eea 
In the last step, we have taken the $\alpha\ll 1$ limit.  The physical interpretation of this solution is that $\phi$ minimizes the electromagnetic energy by (inefficiently) screening the charge inserted by the boundary condition.  This power law can be extremely slow in the small $\alpha$ limit. 
We may also consider the $mr\gg1$ limit, where the gradient term can be neglected in Eq.~\ref{Eq: EoM Nf1} and the mass term can be balanced against the electromagnetic term to find
\bea
\label{Eq: Massive Asymp}
\phi(r)=\begin{cases} \frac{\alpha \theta}{\pi^3 m^2 r^2} &\quad \text{if}\quad q=0\\
2\pi -\frac{\alpha \l 2\pi-\theta \r}{\pi^3 m^2 r^2} &\quad \text{if}\quad q=1 .\\
\end{cases}
\eea

In summary, when $r \ll 1/m$, the solution is constant as the gradient and electromagnetic energies hold $\phi$ constant.  When $r \gg 1/m$, the electromagnetic and mass terms balance each other, leading to a $1/r^2$ fall off.  $r \sim 1/m$ is the transition region where, at least in the $\alpha \ll 1$ limit, $\phi$ falls exponentially until it reaches the $r \gg 1/m$ solution.

\begin{figure}[t]
    \centering
    \includegraphics[width=0.47\textwidth]{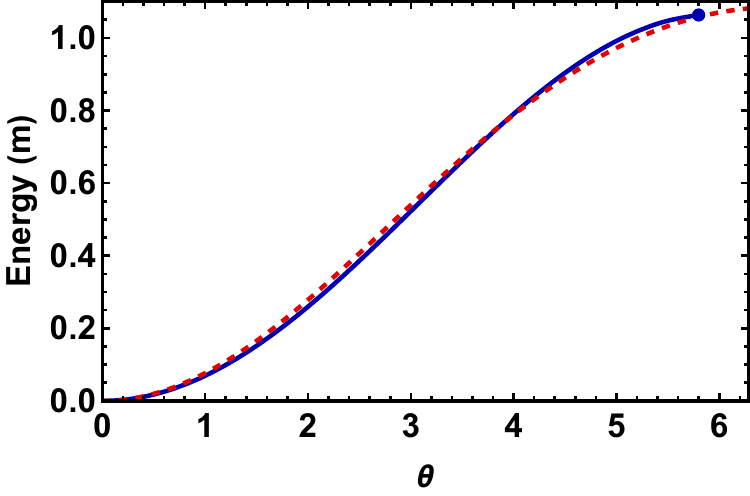}
    \includegraphics[width=0.49\textwidth]{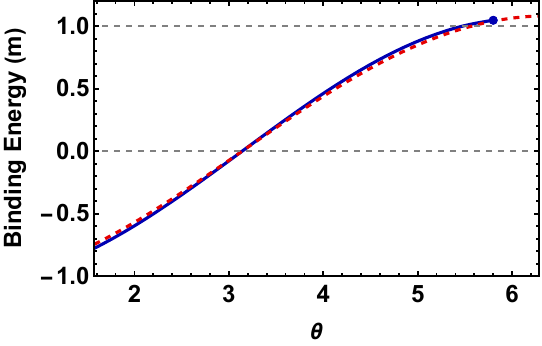}
    \caption{(Left) $\theta$ dependence of the energy of the fermion vacua, $E_0(\theta,\alpha=0.6, m)/m$, and (Right) the binding energy $E_0(\theta,\alpha=0.6, m)/m-E_1(\theta,\alpha-0.6, m)/m$ as a function of $\theta$, both calculated numerically (solid blue lines) and compared against the analytic approximation (dashed red lines) given in Eq.~\ref{Eq: Analytic Approx Energy}.  
    As can be seen in both plots, due to electromagnetic corrections the energy is eventually larger than $m$, indicating that at some point the solution becomes unstable to decay via fermion emission.  At an even higher value of $\theta$, the solution is numerically found to become classically unstable. }
    \label{Fig: theta approx}
\end{figure}

Finally, we are interested in studying the energetics of the solution and the eventual phase transition that occurs as $\theta$ is varied.  Let us first find an approximation for the extra energy of the $\DD_q$ bound state, $E_q(\theta,\alpha,m)$, normalized such that $E_0(\theta=0,\alpha,m)= 0$.  
To start, consider the Hamiltonian for the theory, given below
\bea
\label{Eq: Nf1 Hamiltonian}
H=\frac{1}{4\pi}\int_0^\infty dr\; \frac{1}{2}{\phi'}^2+\frac{1}{2}(m\pi)^2\sin^2\(\frac{\phi}{2}\)+\frac{\alpha}{8\pi r^2}(\phi-\theta)^2.
\eea
This Hamiltonian can be broken into two contributions, the mechanical energy and electromagnetic energy, represented by the first two terms and the last term, respectively.
In the $\alpha = 0$ limit, we can analytically obtain the mechanical energy by multiplying  Eq.~\ref{Eq: EoM Nf1} by $\phi'$ and integrating in from infinity to find
\bea
\label{Eq: equal mass and grad}
(\phi')^2=(m\pi)^2\sin^2(\phi/2).
\eea
Using this, we can integrate the mechanical portion of the Hamiltonian analytically to find
\begin{align}
E_q(\theta,\alpha=0,m)=m\sin^2\(\frac{\pi}{2}Q_{EM}^{tot}(q,\theta)\) ,
\end{align}
which is valid as long as the solution does not contain a soliton at infinity.  We also denote the total charge of the solution $Q_{EM}^{tot}(q,\theta) = q - \theta/2 \pi$.
$\alpha \ne 0$ corrections can be estimated by crudely approximating $\phi$ as a step function with $\phi=\theta$ for $mr<1$ and $\phi=2\pi q$ for $ mr>1$. With this approximation, the electromagnetic contribution to the energy is
\bea
\label{Eq: EM Energy Nf=1}
E_q^{\rm EM}(\theta,\alpha,m)\approx\frac{\alpha}{2}m\(
\frac{Q_{EM}^{tot}(q,\theta)}{2}\)^2 .
\eea
Combining these two contributions to the energy gives an analytic approximation of the energy
\bea
\label{Eq: Analytic Approx Energy}
E_q(\theta,\alpha,m)\approx m\(\sin^2\(\frac{\pi}{2}Q^{tot}_{EM}(q,\theta)\)+\frac{\alpha}{2}\(\frac{Q^{tot}_{EM}(q,\theta)}{2}\)^2\) .
\eea
It is worth noting that we must be careful about the $\alpha =0$ limit, as the proper order of limits is that the size of the monopole goes to zero first, before $\alpha$ is taken to be small.  At the origin, the $\alpha/r^2$ piece is always important.

\begin{figure}[t]
    \centering
    \includegraphics[width=0.49\textwidth]{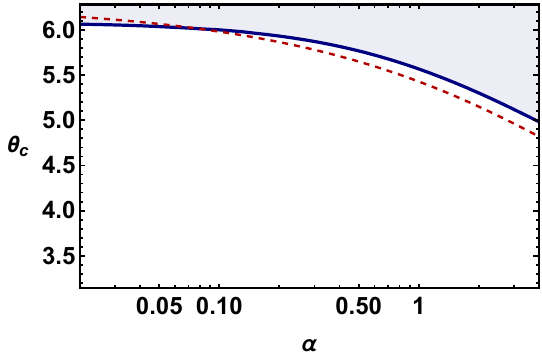}
    \caption{The value of $\theta$ above which a $\DD_0$ bound state becomes quantum mechanically unstable to decay into a $\DD_1$ bound state and a fermion as a function of $\alpha$.  The numerical result for $\theta_c$ is given by the solid line while the analytic approximation in Eq.~\ref{Eq: theta crit analytic Nf1 final} is given by the dashed line.  Symmetry considerations can be used to find the analogous lines for any of the other $\DD_q$ dyons. }
    \label{Fig: thetac}
\end{figure}

We plot the numerical and analytic approximation of $E_q(\theta,\alpha = 0.6,m)$ in Fig.~\ref{Fig: theta approx}.  Two features of this plot are important and generalize well to $N_f > 1$.  The first can be seen from the right-hand plot, which is that whenever $\alpha > 0$, there is a $\theta$ value for which $E_0(\theta)-E_1(\theta) > m$.  The implication is that there is a critical $\theta_c$ above which $\DD_0$ is quantum mechanically unstable to decay via fermion emission and becoming $\DD_1$.  
We can solve for $\theta_c$ using
\bea
\label{Eq: theta crit Nf1}
E_0(\theta_c)-E_1(\theta_c)-m=0 .
\eea
For all $\theta \le\theta_c$, the monopole is stable against decay.  In the $\alpha =0$ and $\infty$ limits, we can analytically find $\theta_c(\alpha=0) = 2 \pi$ and $\theta_c(\alpha = \infty) = \pi$.  Combining Eq.~\ref{Eq: theta crit Nf1} with Eq.~\ref{Eq: Analytic Approx Energy}, we can find an analytic expression for $\theta_c$ valid near $2 \pi$
\bea
\label{Eq: theta crit analytic Nf1 final}
\theta_c \approx 2\pi -\frac{\alpha}{2\pi}\(\sqrt{1+\frac{4\pi^2}{\alpha}}-1\) .
\eea
Fig.~\ref{Fig: thetac} plots $\theta_c$ as a function of $\alpha$.

The next feature to note in Fig.~\ref{Fig: theta approx} is that there is another critical $\theta_{PT}$ at which the solution is numerically found to become classically unstable.  The existence of $\theta_{PT}$ can be most easily seen at $\theta = 2 \pi$.  It is clear that any solution is classically unstable due to $\alpha$ effects, and the only solution is a soliton at infinity.

Finally, we comment on the number of stable solutions as a function of $\theta$.  For $0 < \theta < 2 \pi - \theta_{PT}$, there is just the monopole solution and asymptotic fermions.
For $2 \pi - \theta_{PT} < \theta < 2 \pi - \theta_{c}$, there are two dyonic bound states $\DD_{0}$ and $\DD_1$, though $\DD_1$ is unstable to decay into $\DD_0$ quantum mechanically.  For $2 \pi - \theta_{c} < \theta <\theta_c$, both $\DD_{0}$  and $\DD_1$ are present and the higher-energy state can be viewed as the lower-energy state with a bound-state fermion whose binding energy is of order the mass.  At $\theta > \theta_c$, $\DD_0$ becomes quantum mechanically unstable before finally ceasing to exist as a classical solution when $\theta > \theta_{PT}$.

\section{$N_f=2$}
\label{Sec: Nf2}

In this section, we study the $SO$ and $Sp$ theories for $N_f=2$. In the IR, both theories consist of 4 left-handed Weyl fermions $\ell$, $b$, $\ell^c$ and $b^c$ with mass terms $m_\ell \ell \ell^c + m_b b b^c$.  In the equal mass limit, the IR theory appears to contain an $SU(2)$ flavor symmetry with only $U(1)_{B-L}$ surviving when the masses are unequal.  However, from the UV perspective, the theories are very different, as can be seen in Tab.~\ref{Tab: Nf2 Charges}. Most significantly, $U(1)_{B-L}$ is only a good symmetry for the $Sp(2)$ theory and is explicitly broken in the $SO(2)$ theory. 

\begin{table}[h!]
 \centering
 \begin{tabular}{c c}
 $SO(2)$ Flavor Symmetry & $Sp(2)$ Flavor Symmetry\\
 &\\
\begin{tabular}{c|c|c}
&$SU(2)\; \rar \;U(1)_{EM}$&$U(1)_{B-L}$ \\
\hline
$\begin{pmatrix}
 b \\
 b^c
 \end{pmatrix}$&$\Box \; \rar\; \begin{pmatrix}
 \frac{1}{2} \\
 -\frac{1}{2}
 \end{pmatrix}$&$\begin{matrix}
 1 \\
 -1
 \end{matrix}$ \\
 $\begin{pmatrix}
 \ell \\
 \ell^c
 \end{pmatrix}$&$\Box\; \rar\; \begin{pmatrix}
 \frac{1}{2} \\
 -\frac{1}{2}
 \end{pmatrix}$&$\begin{matrix}
 -1 \\
 1
 \end{matrix}$ \\
\end{tabular}
&
\begin{tabular}{c|c|c}
&$SU(2)\; \rar \;U(1)_{EM}$&$U(1)_{B-L}$ \\
\hline
$ \begin{pmatrix}
 b \\
 \ell^c
 \end{pmatrix}$&$\Box \; \rar\; \begin{pmatrix}
 \frac{1}{2} \\
 -\frac{1}{2}
 \end{pmatrix}$&$\begin{matrix}
 1 \\
 1
 \end{matrix}$ \\
$ \begin{pmatrix}
 \ell \\
 b^c
 \end{pmatrix}$&$\Box\; \rar\; \begin{pmatrix}
 \frac{1}{2} \\
 -\frac{1}{2}
 \end{pmatrix}$&$\begin{matrix}
 -1 \\
 -1
 \end{matrix}$ \\
\end{tabular}
\end{tabular}
\caption{How the baryons and leptons are embeded into two different UV completions labeled by the flavor symmetries present when the baryon and lepton masses are equal.  Note that in the $SO(2)$ theory, $B-L$ is not a good symmetry while in $Sp(2)$ it is a good symmetry\label{Tab: Nf2 Charges}}
\end{table}
After bosonization, the Lagrangian for both theories is 
\begin{align}
\label{Eq: Nf2 Lagrangian}
L=&\frac{1}{4\pi}\int_0^\infty dr\;\frac{1}{2}\(\partial_\mu\phi_\ell\partial^\mu\phi_\ell+\partial_\mu\phi_b\partial^\mu\phi_b\)-\(\frac{\pi m_\ell(r)}{2}\)^2(1-\cos(\phi_\ell))\\ 
\nn-&\(\frac{\pi m_b(r)}{2}\)^2(1-\cos(\phi_b))-\frac{\alpha}{8\pi r^2}\(\phi_\ell+\phi_b-2\theta\)^2 ,
\end{align}
which gives the time-independent equations of motion
\begin{align}
\label{Eq: EoM Nf2}
\partial_r^2\phi_b&=\(\frac{\pi m_b(r)}{2}\)^2\sin(\phi_b)+\frac{\alpha}{4\pi r^2}\(\phi_b+\phi_\ell-2\theta\)\\
\nn
\partial_r^2\phi_\ell&=\(\frac{\pi m_\ell(r)}{2}\)^2\sin(\phi_\ell)+\frac{\alpha}{4\pi r^2}\(\phi_b+\phi_\ell-2\theta\) .
\end{align}
For the $Sp(2)$ theory $m_{\ell,b}(r)=m_{\ell,b}$  and for the $SO(2)$ theory
\bea
\label{Eq: SON mass}
m_i(r)=m_ie^{-\frac{1}{2}K_0(\pi^2 e^{-\gamma}m_i r/4)} \approx m_i\begin{cases}
1 & \text{if}\quad m_ir\gg 1\\
\frac{\pi}{2} \sqrt{\frac{m_i r}{2}} & \text{if}\quad m_ir\ll 1
\end{cases}
\eea
Note that both theories are symmetric under the exchange of the labels $b\longleftrightarrow \ell$, so we can always take $m_b\geq m_\ell$. The boundary conditions at $r=0$ are
\begin{align}
\label{Eq: Nf2 BC}
\text{$SO(2)$:}&\quad \partial_r\phi_\ell=\partial_r\phi_b=0 \\
\text{$Sp(2)$:}&\quad \partial_r\phi_\ell=-\partial_r\phi_b \quad \phi_\ell=\phi_b .
\end{align} 
Since the scalars $\phi_\ell$ and $\phi_b$ are $4 \pi$ periodic, these boundary conditions involve a bit of gauge fixing.
Finite energy forces the boundary conditions $\phi_{b,\ell}(r=\infty)=2\pi n_{b,\ell}$ and $\phi_{b}(0)+\phi_\ell(0)=2\theta$. States that satisfy these boundary conditions carry electromagnetic and $B-L$ charges
\begin{align}
Q_{EM}^{tot}=&\frac{n_b+n_\ell}{2}-\frac{\theta}{2\pi}\\
\nn Q_{B-L}^{tot}=&n_b-n_\ell-\frac{\phi_b(0)-\phi_\ell(0)}{2\pi} .
\end{align}
The total leptonic and baryonic charge in each field can be related to the total electric charge $Q_{EM}^{tot}$ (in units of $e$) and the total $B-L$ charge $Q_{B-L}^{tot}$
\bea
\label{Eq: Total charges}
Q_{b}^{tot}=Q_{EM}^{tot}+\frac{Q_{B-L}^{tot}}{2} \quad Q_{\ell}^{tot}=Q_{EM}^{tot}-\frac{Q_{B-L}^{tot}}{2} .
\eea
Now we come to a very important physical fact. The boundary condition $\phi_\ell(0)=\phi_b(0)$ in the $Sp(2)$ theory forces $Q_{B-L}^{tot}$ to be an integer, reflecting that $U(1)_{B-L}$ is a good symmetry of the theory.
In this case, the total charge stored in each field is fixed by the total electromagnetic charge and the $B-L$ charge of the bound state.  In the $SO(2)$ theory, $U(1)_{B-L}$ is not a good symmetry and so $Q_{B-L}^{tot}$ can be non-integral.  Since $Q_{B-L}^{tot}$ is free to take any value, the bound state solutions in the $SO(2)$ theory can shift the portion of the electric charge stored in each field in order to minimize the energy. As we will see, this extra bit of freedom for the $SO(2)$ theory will have many physical consequences.

\begin{figure}[t]
\centering
 \includegraphics[width=0.7\textwidth]{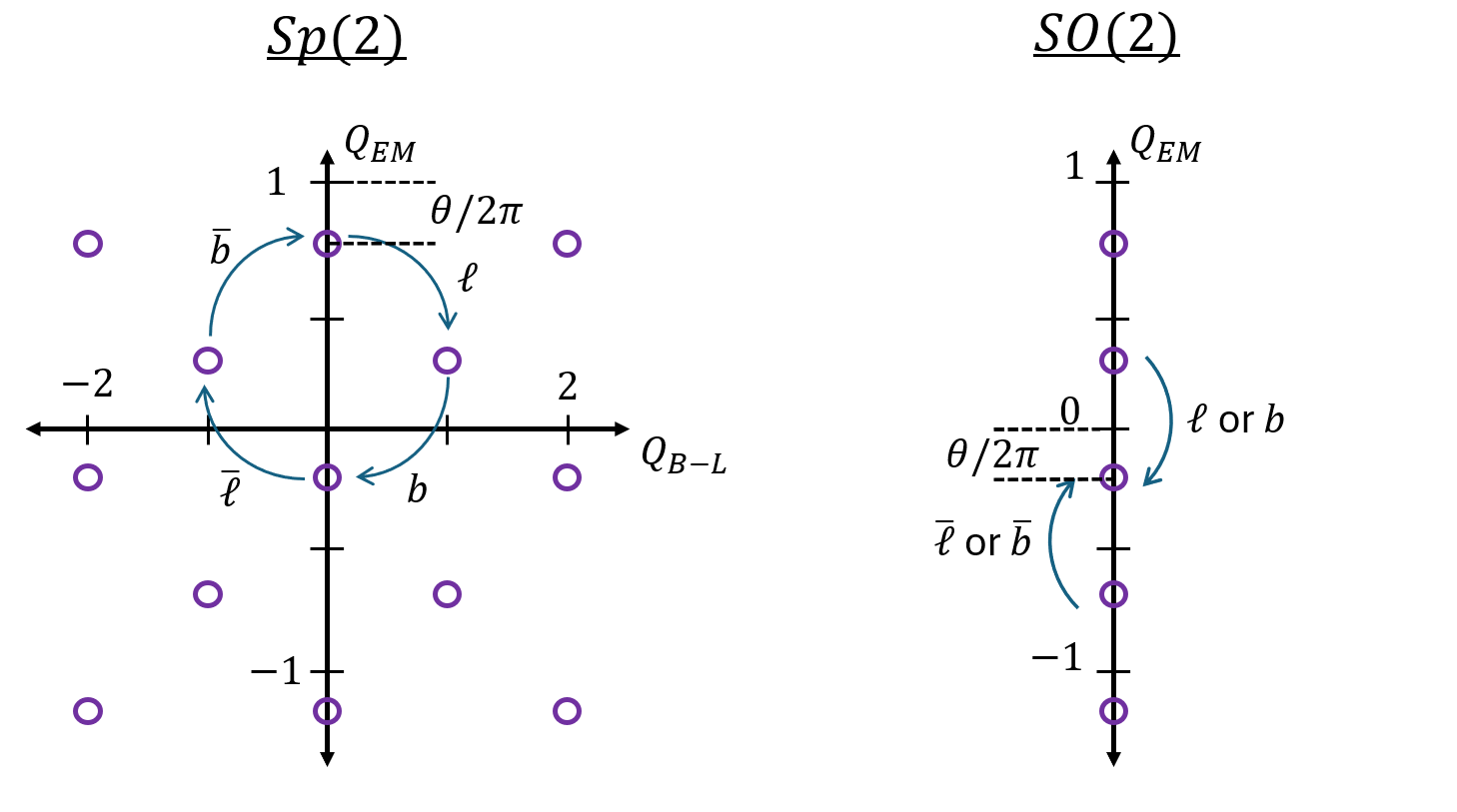}
 \caption{\label{Fig: States Sketch} A map of the various magnetically and electrically charged states in the $Sp(2)$ and $SO(2)$ theories. The arrows indicate the particles emitted by one state to decay into another. Note in $SO(2)$, the states can decay by either emitting a (anti)lepton or (anti)baryon but if $m_\ell<m_b$, it will energetically prefer to decay via a lepton.}
\end{figure}

We now move on to the issue of labeling/counting all of the dyonic bound states in the theory.
In analogy with our $N_f=1$ case, the first expectation is that there is one dyonic bound state (stable or not) for every combination of $n_\ell$ and $n_b$.   While this is true for the $Sp(2)$ theory, many of these states are equivalent in the $SO(2)$ theory. To see this, consider the following field redefinitions $S_\ell(n)$ and $S_b(n)$
\begin{align}
\label{Eq: Shift Syms}
\text{$S_b(n)$:} &\quad \phi_\ell\rar \phi_\ell\quad \phi_{b}\rar\phi_{b}+2\pi n \quad \theta\rar \theta +\pi n\\ \nn\text{$S_\ell(n)$:} &\quad \phi_{\ell}\rar\phi_{\ell}+2\pi n \quad \phi_b\rar \phi_b\quad \theta\rar \theta+\pi n .
\end{align}
The $S_\ell(1)$ field redefinition in the fermionic language corresponds to $\psi_\ell \rightarrow - \psi_\ell$ and $\theta \rar \theta + \pi$, with similar expressions holding for $S_b(1)$.
The Lagrangian in Eq.~\ref{Eq: Nf2 Lagrangian} and the $SO(2)$ boundary conditions are invariant under both of these field redefinitions.  These can be used to equate dyonic bound state solutions in a theory with one value of $\theta$ to dyons in a theory with another value of $\theta$. If one considers the redefinition $S_-(n)\equiv S_b(n)S_\ell(-n)$, one finds that it relates dyons in the same theory, as this redefinition sends $\theta\rar \theta$. One finds an equivalence between states 
$(n_b,n_\ell)\cong (n_b+1,n_\ell-1)$. This collapses the dyonic states into a set of equivalence classes which may be indexed by a single quantum number $q=n_b+n_\ell$. Thus dyonic solutions are uniquely determined by their electric charge $Q_{EM}^{tot}=q/2-\theta/2\pi$. We label the dyonic bound states in $SO(2)$ as $\DD_q$.  The landscape of the states is shown in Fig.~\ref{Fig: States Sketch}. 
Once this equivalence has been made, one can use either $S_\ell(q)$ or $S_{b}(q)$ to relate the dyons as $\theta$ changes by a full period of $q \pi$ to be
\bea
\label{Eq: SO(2) symmetry relation}
\DD_q(\theta)=\DD_0(\theta-q\pi)
\eea
with the associated equality of energies
\bea
\label{Eq: SO2 Energy Relations}
\quad E_{q}(\theta)=E_{0}(\theta-q\pi).
\eea

A similar analysis can be performed on the $Sp(2)$ theory. Here, the boundary conditions do not allow the field redefinitions $S_\ell(n)$ and $S_b(n)$ individually, but only the combination $S_+(n) \equiv S_b(n)S_\ell(n)$. This means that there is no equivalence between states of different $n_\ell$ and $n_b$ in this theory and each combination specifies a unique state. Rather than using $n_\ell$ and $n_b$ to index our states, it will be more useful to index them as $\DD_{q,n_{B-L}}$, with $n_{B-L} = n_b - n_\ell$, as this will reflect more clearly the electric and $B-L$ charges of the specific states. Then we can use $S(n)$ to derive an equivalence between states in theories of different $\theta$ and thus their energies
\bea
\label{Eq: Sp(2) symmetry relation}
\DD_{q,n_{B-L}}(\theta)=\DD_{q-2n,n_{B-L}}(\theta-2\pi n) .
\eea
This implies that the energies of these states are the same in the two theories
\bea
\label{Eq: Sp2 Energy Relations}
\quad E_{q,n_{B-L}}(\theta)=E_{q-2n,n_{B-L}}(\theta-2n\pi) .
\eea
This equivalence of states in different theories is most easily visualized from the diagrams in Fig.~\ref{Fig: States Sketch} where one can see that by shifting $\theta$ a certain amount, the states match up with states of either higher or lower electric charge.

\subsection{Dyonic bound state solutions}
 
\begin{figure}[t]
\centering
 \includegraphics[width=0.49\textwidth]{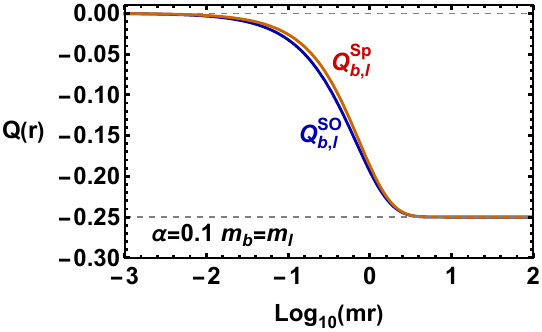}
 \includegraphics[width=0.49\textwidth]
 {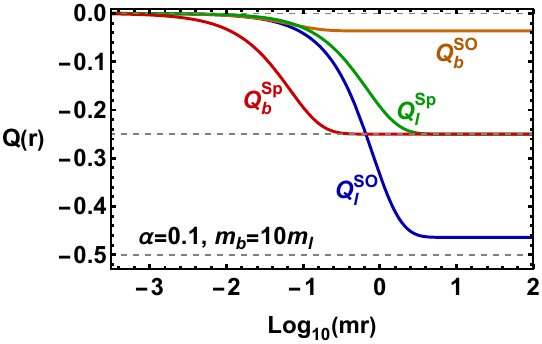}
 \includegraphics[width=0.49\textwidth]{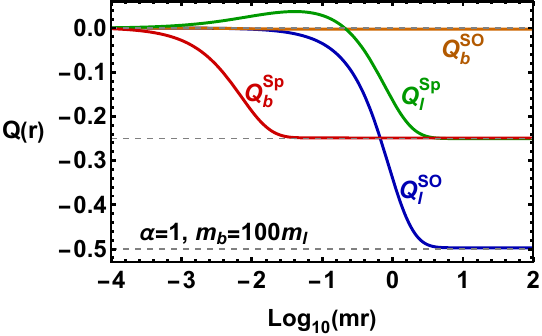}
 \includegraphics[width=0.49\textwidth]
 {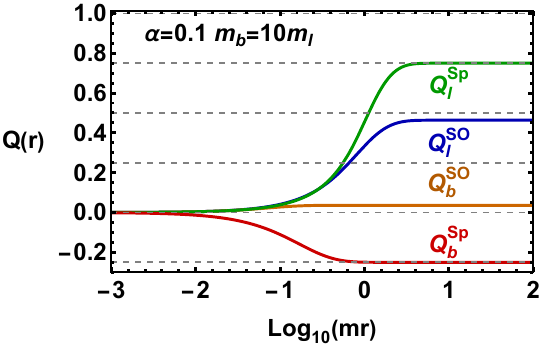} 
 \caption{\label{Fig: Nf2 Dyons} A few examples of what the $\theta = \pi/2$ dyonic bound states look like in both the $SO$ and $Sp$ theories for various mass ratios and values of $\alpha$.  We plot the baryon and lepton number enclosed in a radius $r$ (multiply by $1/2$ for the electric charge) as a function of $r$.  In the top row and bottom left, we show $\DD_0$ and $\DD_{0,0}$, while in the bottom right, we show $\DD_1$ and $\DD_{1,-1}$.}
\end{figure}

\begin{figure}[t]
\centering
 \includegraphics[width=0.49\textwidth]{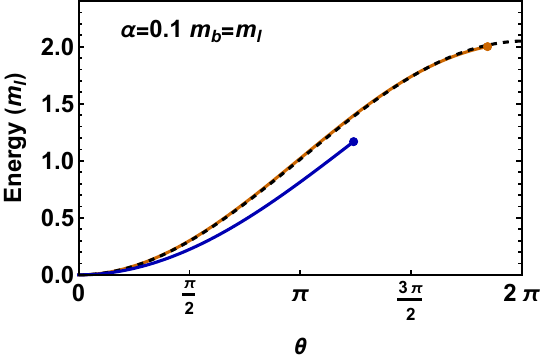}
 \includegraphics[width=0.49\textwidth]
 {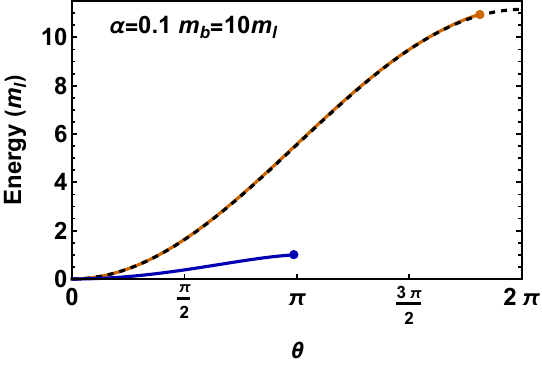}
 \includegraphics[width=0.49\textwidth]{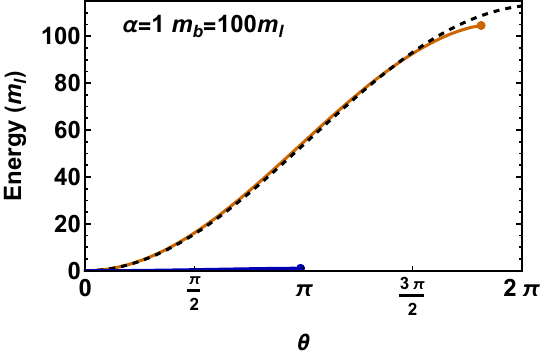}
 \includegraphics[width=0.49\textwidth]
 {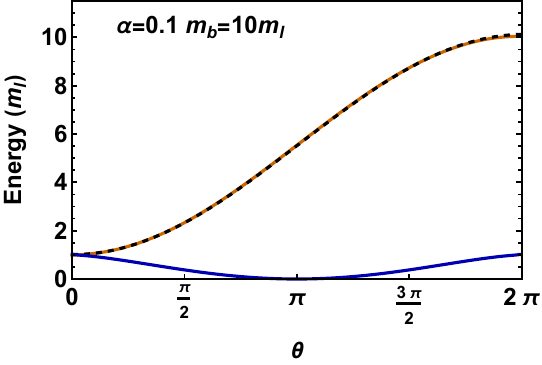}
 \caption{\label{Fig: Nf2 Dyon Energy} A plot of the energies of the states in Fig.~\ref{Fig: Nf2 Dyons} as a function of $\theta$.  The blue lines show the $SO(2)$ theory, the orange lines show the $Sp(2)$ theory, and the black dashed line shows the analytic approximation for the $Sp(2)$ theory given in Eq.~\ref{Eq: Sp2 Analytic Energy}.  It is clear that the $Sp$ theory's energy dependence is dictated by the heavier fermion, while the $SO$ theory's energy dependence is dictated by the lighter fermion.  As before the top row and bottom left show the $\theta = \pi/2$ dyonic bound states $\DD_0$ and $\DD_{0,0}$ while the bottom right shows the $\theta = \pi/2$ dyonic bound states $\DD_1$ and $\DD_{1,-1}$.}
\end{figure}

Fig.~\ref{Fig: Nf2 Dyons} shows a few examples of the dyonic bound-state solutions in the $SO$ and $Sp$ theories.  In the equal-mass case (top left), the $SO$ and $Sp$ theories are essentially identical as everything is equally distributed between the lepton and baryon.  The only difference is due to the mass term having $r$ dependence in the case of the $SO$ theory.
Once the masses are unequal, there are some striking contrasts between the two theories due to the importance of their respective boundary conditions.  
As discussed before, $Sp(2)$ states have a fixed amount of charge in each field, while the $SO(2)$ states are free to give different charges to each field, resulting in them having more freedom to minimize their energy.  

Once $m_b\gg m_\ell$, the $SO(2)$ states prefer to allocate very little charge to the baryon field (solid orange lines) since doing so costs $\OO(m_b)$ energy, as opposed to allocating it to the lepton(solid blue lines) which costs $\OO(m_\ell)$ energy. The $Sp(2)$ states, on the other hand, are forced to split the charge according to Eq.~\ref{Eq: Total charges}, and thus the state has no choice but to acquire $\OO(m_b)$ energies. This can be seen from the energy plots in Fig.~\ref{Fig: Nf2 Dyon Energy} where the energy of the $Sp(2)$ states are $\OO(m_b)$ and the energies of the $SO(2)$ states are $\OO(m_\ell)$. 

There is one more interesting piece of behavior in the $m_b\gg m_\ell$ limit of the $Sp(2)$ theory.  From the bottom-left plot in Fig.~\ref{Fig: Nf2 Dyons}, we see the lepton attempt to screen the baryon's negative charge at $r\approx m_b^{-1}$ by slowly depositing positive charge. This screening can be quantified by the approximate analytic solution in Eq.~\ref{Eq: massless asymp}. In this intermediate region $m_b^{-1}<r<m_{\ell}^{-1}$, we can ignore the mass term and apply the condition that $\phi_{\ell}(r=m_b^{-1})=\theta$ to find the approximate solution 
\bea
\label{Eq: Intermediate phi ell}
Q_\ell(r)\approx -Q_b^{tot}(1-(m_br)^{-\beta}).
\eea
Physically, the lepton is attempting to screen the charge deposited by the baryon to lower the electromagnetic energy. Over this intermediate region, the lepton deposits an electromagnetic charge, $Q_{\ell,screen}^{tot}/2$, given by 
\bea
\label{Eq: Qscreen}
Q_{\ell,screen}^{tot}=Q_b^{tot}((m_b/m_\ell)^{-\beta}-1)\approx\frac{\alpha Q_b^{tot}}{4\pi}\ln\(\frac{m_\ell}{m_b}\)
\eea
where we have taken $\alpha/4\pi\ll1$ in this last approximation. This screening is extremely weak for small $\alpha$, which can be seen as a consequence of the equations of motion. Since we are ignoring the lepton mass term at these small distances $r \approx 1/m_b$ from the core, $\partial_r^2\phi_\ell\sim \OO(\alpha m_b^2/4\pi) $ which limits the rate at which the lepton can deposit charge. Comparatively, the baryon can deposit its charge quickly with $\partial_r^2\phi_b\sim \OO(m_b^2)$.

Now let us consider the energy of these states, shown in Fig.~\ref{Fig: Nf2 Dyon Energy}. Although the two dyons in the top-left plot in Fig.~\ref{Fig: Nf2 Dyons} differ very little, the corresponding energies in Fig.~\ref{Fig: Nf2 Dyon Energy} differ by an $\OO(1)$ factor.  This is because, although the $SO(2)$ theory's radially-dependent mass term does not influence the charge distribution of the state significantly, it does significantly change the energy since the energy from the mass term in the Hamiltonian is suppressed when $mr<1$. Unfortunately, this radial dependence prevents us from using the analytic tricks used in Sec.~\ref{Sec: Nf1} to compute the energy.  As a result, any analytic approximation is only valid at the $\OO(1)$ level for the $SO(2)$ theory.  We do not have this issue for the $Sp(2)$ theories, and we can derive an analytic approximation using many of the same techniques as in Sec.~\ref{Sec: Nf1}. We find 
\begin{align}
\label{Eq: Sp2 Analytic Energy}
E_{q,n_{B-L}}(\theta,\alpha,m_b,m_\ell) \approx \, &m_b\sin^2\(\frac{\pi}{2}Q_b^{tot}\)+m_\ell\sin^2\(\frac{\pi}{2}(Q_\ell^{tot} - Q_{\ell,screen}^{tot})\)\\ \nn &+m_\ell\frac{\alpha}{2}(Q_{EM}^{tot})^2+E_{mid}(\theta,\alpha,m_b,m_\ell).
\end{align}
There are two main new considerations. First, due to the aforementioned screening, the lepton deposits a little extra charge at $r=m_\ell^{-1}$ in addition to its total charge. Second, there is some additional mechanical and electromagnetic energy in the intermediate region $m_b^{-1}<r<m_{\ell}^{-1}$ (denoted $E_{mid}$), which was not present in our $N_f=1$ example. $E_{mid}$ can be computed in this region using Eq.~\ref{Eq: Intermediate phi ell}. Inserting this solution into the Hamiltonian and ignoring the mass term in the intermediate region, we find
\begin{align}
\label{Eq: Emid}
E_{mid}&=\int_{m_b^{-1}}^{m_\ell^{-1}}dr \frac{\pi{Q_\ell'}^2(r)}{2}+\frac{\alpha}{8 r^2}(Q_\ell(r)+Q_b^{tot})^2\\ \nn
&=\frac{\pi}{2} \frac{\beta^2+\alpha/4\pi}{
1+2\beta}(Q_b^{tot})^2\(m_b-m_\ell\(\frac{m_\ell}{m_b}\)^{2\beta}\)\approx\frac{\alpha}{2}\(\frac{Q_b^{tot}}{2}\)^2(m_b-m_\ell) ,
\end{align}
where in the last line of each we have taken the $\alpha/4\pi\ll1$ limit. The analytic approximation given in Eq.~\ref{Eq: Sp2 Analytic Energy} with Eq.~\ref{Eq: Qscreen}-~\ref{Eq: Emid} is plotted alongside the numerical result in Fig.~\ref{Fig: Nf2 Dyon Energy}, showing good agreement.

\subsection{Ground State and Stability}

We are now in the position to ask about the stability of these bound states. First, we can ask which state is the ground state at a particular point in parameter space by finding the state of lowest energy. For the $SO(2)$ theory, it is easy to see from Eq.~\ref{Eq: SO2 Energy Relations} that the ground state is always the state of minimal electric charge, as one might have naively expected. 

\begin{figure}[t]
\centering
 \includegraphics[width=0.49\textwidth]{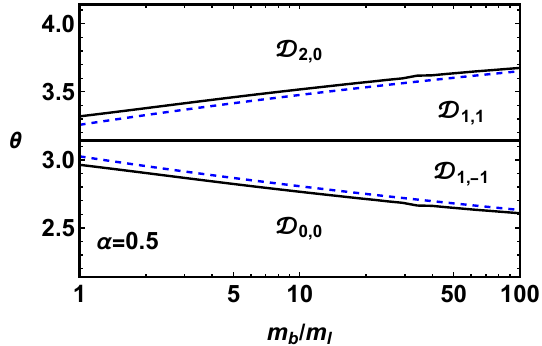}
 \includegraphics[width=0.49\textwidth]
 {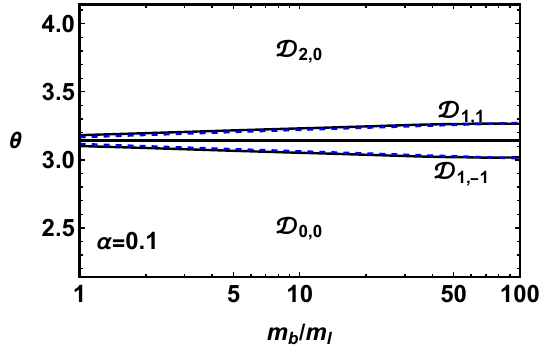}
 \includegraphics[width=0.51\textwidth]{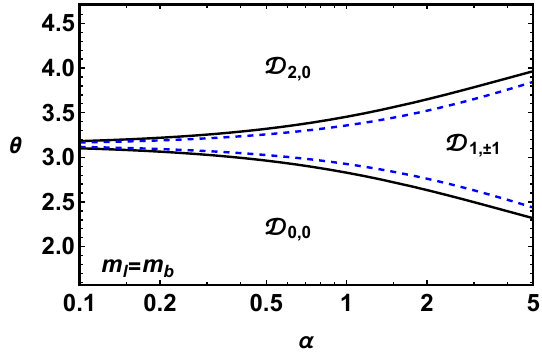} 
 \caption{\label{Fig: Sp2 Ground State} A plot of the regions of parameter space where the different $Sp(2)$ bound states are the ground state. The top two plots show the ground state as a function of $\theta$ and $m_b/m_\ell$ for different values of $\alpha$, while the bottom plot shows the ground state for equal masses as a function of $\theta$ and $\alpha$. The blue dashed lines represent the analytic approximation, while the the solid black lines represent the numerical result. }
\end{figure}

The situation is not so simple for $Sp(2)$, as can be seen in Fig.~\ref{Fig: Sp2 Ground State} where we show the ground state as a function of mass difference and coupling using both our numerical and analytic computations for the energy. These plots can be roughly understood by considering the energy of the $\DD_{1,-1}$ state and the $\DD_{0,0}$ state when $\theta \lesssim \pi$.  From Eq.~\ref{Eq: Sp2 Analytic Energy}, we can see that in the $\alpha\rar 0$ limit, the energies of the two states are 
\bea
\label{Eq: Simple energy Sp2 Ground}
E_{0,0}(\theta,\alpha=0,m_b=m_\ell)=2m\sin^2(\theta/4)\quad E_{1,-1}(\theta,\alpha=0,m_b=m_\ell)=m .
\eea
In this limit, $\DD_{0,0}$ will clearly always be the ground state for $\theta<\pi$. Turning on $\alpha$ gives $\DD_{0,0}$ some extra electric energy whereas it does not give as much to $\DD_{1,-1}$, since $\DD_{1,-1}$ is approximately electronically neutral around $\theta=\pi$. So as we increase $\alpha$ we expect to see more and more area in parameter space where $\DD_{1,-1}$ is the ground state, which is exactly what we see in Fig.~\ref{Fig: Sp2 Ground State}.  Meanwhile, the flip between $D_{1,\pm1}$ at $\theta = \pi$ is easily seen by minimizing the charge held by the heavier baryon as opposed to the lighter lepton.

Finally, we can ask which of these bound states are stable. One quick way to eliminate many states from the list of stable candidates is to note that if $|Q_\ell^{tot}|\geq 1$ or $|Q_b^{tot}| \geq 1$, then the fermion number in at least one of the fields is $\geq1$ and the energy is minimized by placing fermions at spatial infinity, indicating that we are not considering a bound state.  As we are interested in dyonic bound states, this limits us to considering only states with electric charge $|Q_{EM}^{tot}| < 1$ and for $Sp(2)$ $|Q_{B-L}^{tot}|\leq 1$.  As such, we are considering at most 3 states in $SO(2)$ and 4 states in $Sp(2)$, as can be seen from Fig.~\ref{Fig: States Sketch}. If the difference between the energies of any two of these states is greater than the energy of a soliton, then the more energetic dyon can decay into the lower energy dyon through the emission of a soliton, as shown in Fig.~\ref{Fig: States Sketch}. This decay must conserve electric charge and, for the $Sp(2)$ theory, $B-L$ charge. The different charge conservation considerations and the different relations between the energies for the two theories in Eqs.~\ref{Eq: SO2 Energy Relations} and~\ref{Eq: Sp2 Energy Relations} cause the landscape of stable dyonic states to be very different between the two theories.

\begin{figure}[t]
\centering
 \includegraphics[width=0.49\textwidth]{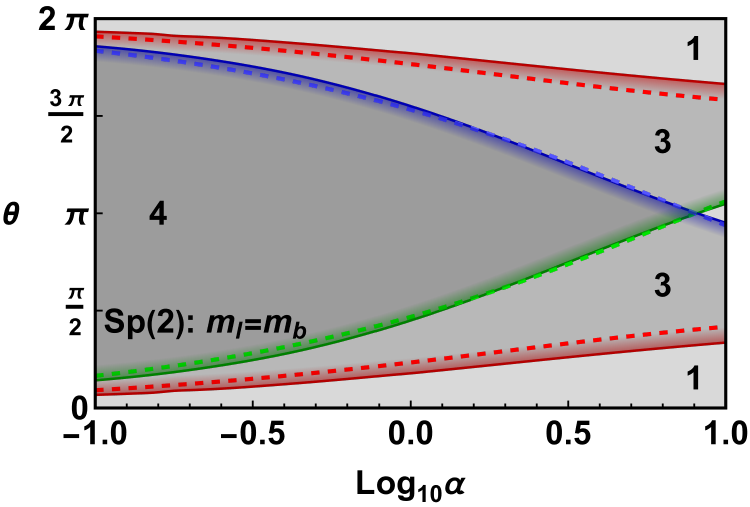}
 \includegraphics[width=0.49\textwidth]
 {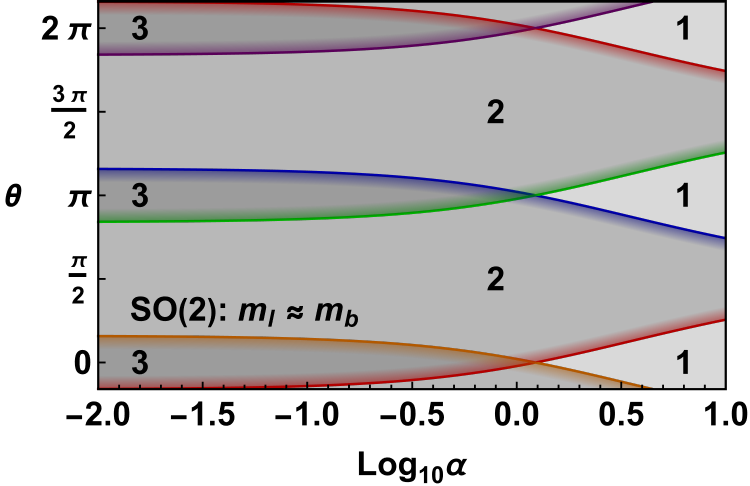} 
 \caption{\label{Fig: Eq Mass Stability} A plot of the various stable bound states as a function of $\theta$ and $\alpha$ for the two theories with the $Sp(2)$ theory ($m_b=m_\ell$) on the left and the $SO(2)$ theory ($m_b=m_\ell + \epsilon$) on the right.  In both plots, the number of stable states in each region is labeled with the darkness of the gray shading representing the number of stable states with the lightest shading being 1 stable state (the ground state), and the darkest being 4 stable states.   The different colors represent the stability regions of different states of the $SO$ ($Sp$) theory with blue corresponding to $\DD_0$ ($\DD_{0,0}$), red corresponding to $\DD_1$ ($\DD_{1,\pm1}$), and green indicating $\DD_2$ ($\DD_{2,0}$).  For the $SO(2)$ plot, the orange and purple correspond to the $\DD_{-1}$ and $\DD_3$ states, respectively. The colored shading around each line points in the direction where the state is stable. The dashed lines on the $Sp(2)$ plot indicate the results from the analytic approximation for the energies.  The solid lines are the numerical results. } 
\end{figure}

\begin{figure}[t]
\centering
 \includegraphics[width=0.49\textwidth]{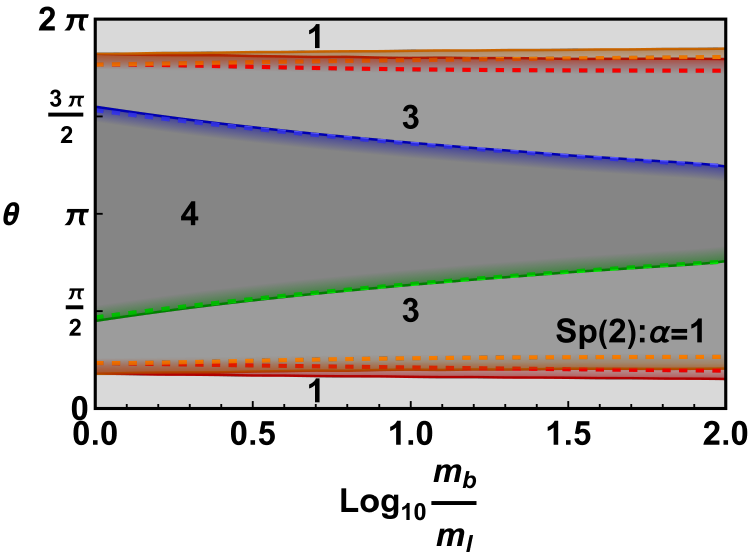}
 \includegraphics[width=0.49\textwidth]
 {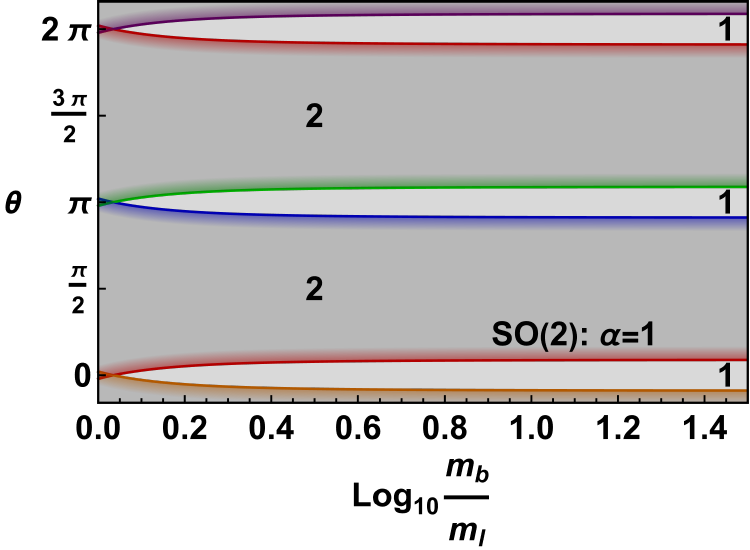} 
 \includegraphics[width=0.49\textwidth]{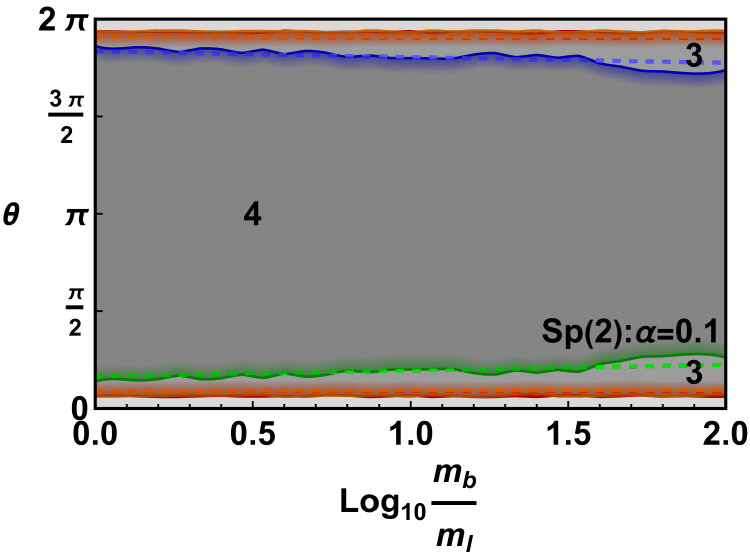}
 \includegraphics[width=0.49\textwidth]
 {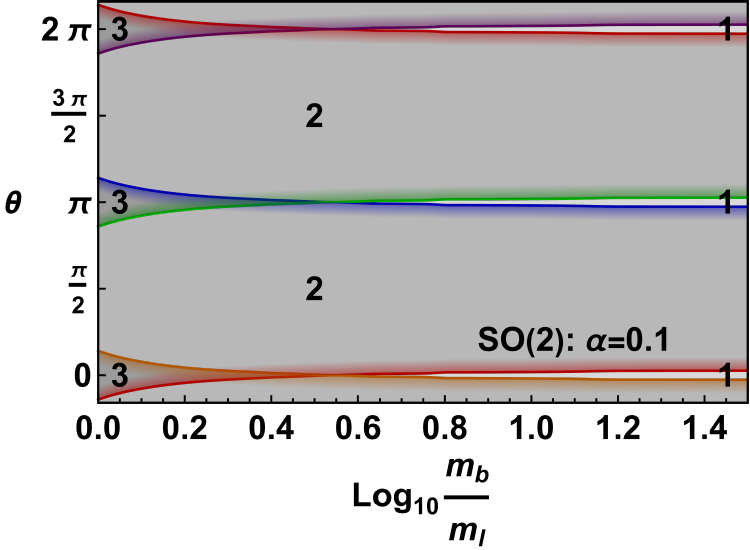} 
 \caption{\label{Fig: alpha Stability} A plot of the various stable bound states as a function of $\theta$ and $m_b/m_{\ell}$ for $\alpha=1$ and $0.1$, with the $Sp(2)$ theory on the left and the $SO(2)$ theory on the right. The color scheme is the same as that in Fig.~\ref{Fig: Eq Mass Stability}, only now with the red and orange lines differentiating between the stability of the $\DD_{1,-1}$ and $\DD_{1,1}$ states, respectively.} 
\end{figure}

Figs.~\ref{Fig: Eq Mass Stability} and~\ref{Fig: alpha Stability} show these regions of stability for the various relevant states in the two theories. These plots show some very interesting features. While for $Sp(2)$ the only stable state at $\theta=0$ is the monopole state $\DD_{0,0}$, for $SO(2)$ there can be potentially 3 stable dyonic bound states, $\DD_{\pm 1}$ and $\DD_{0}$, for sufficiently small coupling $\alpha$ and mass ratio $m_b/m_\ell$. The difference between these two theories can be understood by considering the $B-L$ symmetry. As discussed before, the boundary conditions in the $Sp(2)$ theory fix the amount of charge that must be held by each field. In particular, for $\DD_{1,-1}$ at $\theta=0$, all of the electric charge $e/2$ is carried by the lepton. But a lepton state with charge $e/2$ has a minimum energy configuration of a soliton at infinity, and so there is no bound state.  The same argument cannot be made in the $SO(2)$ theory, where no such symmetry forces the entirety of the charge into one field.  The two fields split the charge between themselves.  They share the energetic burden and have total energy less than a soliton at infinity.

Finally, we comment on the $SO(2)$ symmetric limit where $m_b = m_l$.  Away from this limit, the $\theta = \pi$ dyon decays into the $\theta = 0$ monopole by emitting a baryon or lepton.  In the equal mass limit, conservation of the flavor symmetry prevents this decay as both the $\theta = \pi$ dyon and monopole are flavor neutral and thus the $\theta = \pi$ dyon cannot be viewed as a monopole plus light fermion bound state.  If there is a heavier fermion with no flavor symmetries, then the $\theta = \pi$ dyon can be viewed as a tightly bound state of this heavier fermion.  However as this interpretation and stability analysis requires introducing a heavier fermion, we stay away from the exact $SO(2)$ conserving limit when discussing stability.

\section{$N_f\geq4$ }
\label{Sec: Nfgeq4}

We now move on to consider $SO(N_f)$ and $Sp(N_f)$ theories for $N_f>2$. Both theories consist of $N_f/2$ copies of the $N_f=2$ theory and so we label the fields $\phi_{\ell,i}$ and $\phi_{b,i}$ for families $i=1,2,...,N_f/2$. Symmetry considerations such as the ones in Eq.~\ref{Eq: Shift Syms} show that the states in the $SO(N_f)$ theory are indexed only by $q$, while the $Sp(N_f)$ bound states are indexed by $q$ and $n_{B-L,i}$, for the $i^\text{th}$ $B-L$-type charges of each family. We denote bound states in the $SO(N_f)$ theory as $\DD_q$ and in the $Sp(N_f)$ theory as $\DD_{q,\vec n_{B-L}}$, where $\vec n_{B-L}\in \ZZ^{N_f/2}$ is a $N_f/2$ sized vector of integers. The $Q_{B-L,i}^{tot}$ charge and an electric charge in each family $Q_{EM,i}^{tot}$ for any bound state is 
\begin{align}
\label{Eq: Nf total charges}
Q_{B-L,i}^{tot}=Q_{b,i}^{tot}-Q_{\ell,i}^{tot} \\ \nn
Q_{EM,i}^{tot}=\frac{Q_{b,i}^{tot}+Q_{\ell,i}^{tot}}{2} .
\end{align}
Note that neither theory has a condition that fixes the individual electric charge, $Q_{EM,i}^{tot}$, in each family. The dyons are free to split up the electric charge between the families in any way to minimize the energy. Due to the large number of fields to keep track of, mapping out the entire landscape of stable states as we did for $N_f=2$ is much more complicated. As such, we will  focus on two particular quantities: $\theta_c$, the critical angle at which the $\DD_0$ ($\DD_{0,\vec 0}$) monopole state becomes unstable, and $N_{st}$, the number of stable states at $\theta=0$. We will do this for both theories in the limit of equal masses and briefly discuss the limit where we have two classes of fermions with either a heavy mass $M$ or a light mass $m$.

\subsection{Equal Mass Fermions}

Let us start by considering the case where all fermions have equal mass.  We are first interested in finding $\theta_c$, the angle at which the $\DD_0$ ($\DD_{0,\vec 0}$) state becomes unstable. To start, we will need to find the energy of the $\DD_0$ ($\DD_{0,\vec 0}$) state, which we will call $E_0^{N_f}$, and the energy of the most favorable state for it to decay into, $E_{out}^{N_f}$. $\theta_c$ can then be defined as 
\bea
\label{Eq: thetac def}
E_{0}^{N_f}(\alpha,\theta_c)-E_{out}^{N_f}(\alpha,\theta_c)-m=0.
\eea
In the minimum energy configuration, the total electric charge is evenly distributed between the $N_f$ fields in the $q=0$ state, and thus $\phi_1=\phi_2=...=\phi_{N_f}$. Since all the fields are identical, the Hamiltonian reduces to $N_f/2$ copies of the $N_f=2$ Hamiltonian, with slightly modified values for $\theta$ and $\alpha$
\begin{align}
H&=\frac{N_f}{2}\frac{1}{4 \pi} \int_0^\infty dr\frac{{\phi_{b}'}^2}{2}+\frac{{\phi_{\ell}'}^2}{2}+\(\frac{\pi m(r)}{2}\)^2\(2-\cos(\phi_b))-\cos(\phi_\ell)\) \nn\\
&+\frac{\alpha N_f/2}{8\pi r^2}\(\phi_\ell+\phi_b-\frac{4\theta}{N_f}\)^2 .
\end{align}
$E_0^{N_f}$ can be directly related to $E_0^{N_f =2}$ in an $N_f=2$ theory with effective coupling $\alpha_{eff}=\alpha N_f/2$ and effective $\theta_{eff}=2\theta/N_f$ through
\bea 
\label{Eq: E0000000=E00}
E^{N_f}_0(\alpha,\theta)=\frac{N_f}{2}E^{N_f=2}_{0}(\alpha N_f/2,2\theta/N_f).
\eea
This is true in both $SO(N_f)$ and $Sp(N_f)$ theories where we have suppressed the $\vec n_{B-L}= \vec 0$ index for $Sp(N_f)$. 

\begin{figure}[t]
 \centering
 \includegraphics[width=0.49\textwidth]{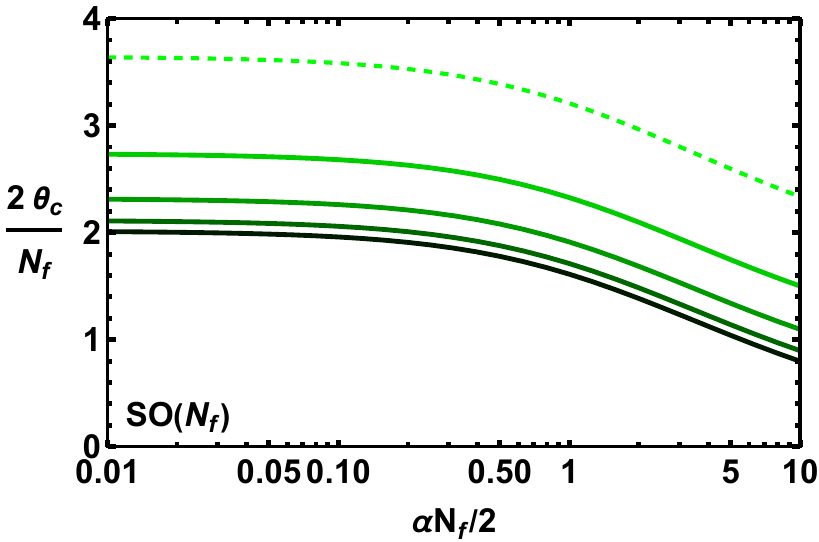}
 \includegraphics[width=0.49\textwidth]{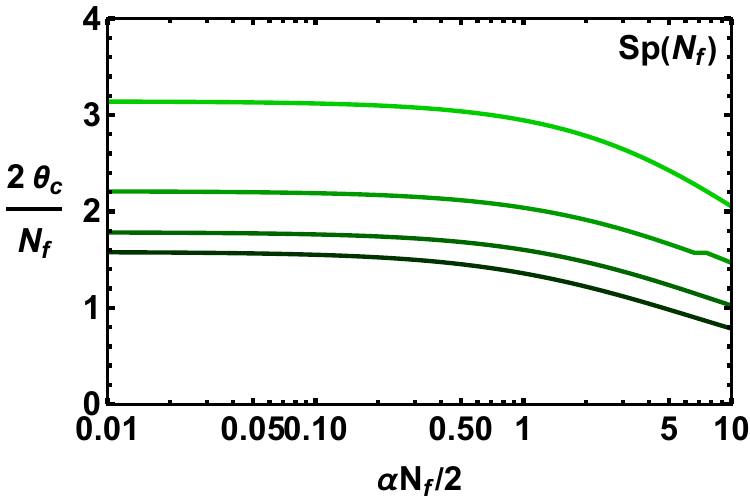}
 \includegraphics[width=0.49\textwidth]{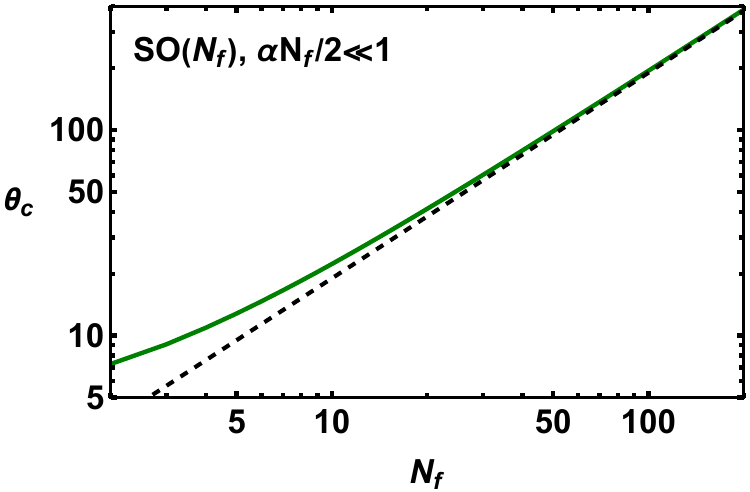}
 \includegraphics[width=0.49\textwidth]{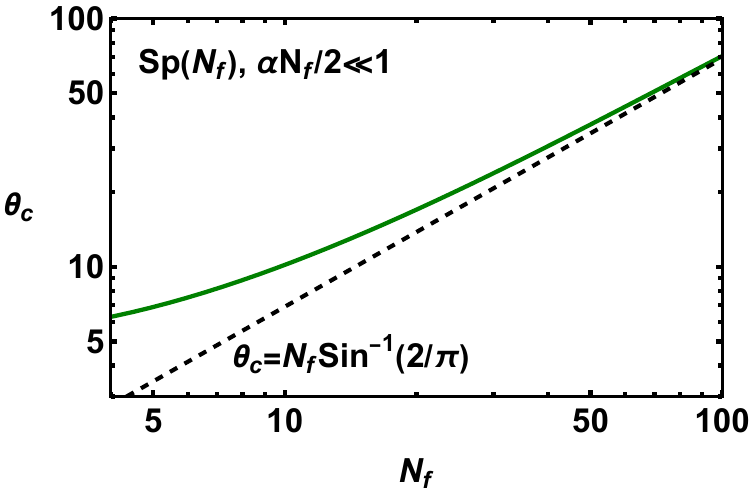}
 \caption{\label{Fig: thetac Nf} The top two plots show $2\theta_c/N_{f}$ as a function of $\alpha N_f/2$ for the two theories. We show the results for $N_f=4,8,16$, and $32$ with the darkest solid green line indicating $N_f=32$ and the lightest solid green line indicating $N_f=4$, with the green getting darker as $N_f$ gets larger (the dotted line indicates $N_f=2$). The bottom two plots show $\theta_c$ as a function of $N_f$ in the limit $\alpha N_f\ll1$, with the solid line being the numerical result. For the $SO(N_f)$ theory, the linear fit $\theta_c\approx 1.9 N_f$ is shown by the dotted line, while for the $Sp(N_f)$ theory the dotted line is instead the analytic approximation $\theta_c\approx \sin^{-1}(2/\pi)N_f$.}
\end{figure}

Next, we find the most favorable state for the $\DD_0$ and $\DD_{0,\vec 0}$ states to decay into. For the $SO(N_f)$ theory, this will naturally be the $\DD_1$ states~\footnote{As mentioned before, only when the masses are epsilonically different are $\DD_1$ and $\DD_0$ related by emission of a single fermion.  If the masses are exactly equal, then $\DD_0$ must emit two fermions in a flavor neutral state to decay into $\DD_2$.  In the large $N_f$ limit, both of these processes give the same value for $\theta_c$.}.  We use a relation analogous to those in Eq.~\ref{Eq: SO(2) symmetry relation} to relate the energy of the $\DD_1$ dyon to that of $\DD_0$ in a theory with $\theta\rar\theta-\pi$ as follows :
\bea
\label{Eq: SON E00000000}
E^{N_f}_1(\alpha,\theta)=E^{N_f}_{0}(\alpha,\theta-\pi)=\frac{N_f}{2}E^{N_f=2}_{0}(\alpha N_f/2,2(\theta-\pi)/N_f) .
\eea
Eq.~\ref{Eq: thetac def}, Eq.~\ref{Eq: E0000000=E00} and Eq.~\ref{Eq: SON E00000000} can be used in conjunction with the numerical energies we computed for $N_f=2$ to find $\theta_c$, which is shown in Fig.~\ref{Fig: thetac Nf}. We also perform a fit to a power law for $N_f\gg1$, $\alpha N_f/2\ll1$ from which we find $\theta_c \approx 1.9 N_f$.

The situation in the $Sp(N_f)$ theory is a bit more complicated. To see why, consider the $\DD_{1,(1,0,...,0)}$ state. This carries electric charge $Q_{EM}^{tot}=1/2-\theta/2\pi$, and $B-L$ type charges $Q_{B-L,1}^{tot}=1$, and $Q_{B-L,i>1}^{tot}=0$. The $i>1$ families minimize energy by sharing equal electric charge; however, because of the asymmetry between the $i=1$ family and the others, there is no reason to suppose that the charge in the $i=1$ is equal to the charge in the other families.  The energy for this state can be found by generalizing Eq.~\ref{Eq: Sp2 Analytic Energy} to arbitrary $N_f$
\begin{align}
\label{Eq: D100000 Energy}
E^{N_f}_{1,,(1,0,...,0)}(\alpha,\theta) \approx &m\sin^2\(\frac{\pi}{2}(Q_{EM,1}^{tot}+1/2)\)+m\sin^2\(\frac{\pi}{2}(Q_{EM,1}^{tot}-1/2)\)+\\
\nn &+(N_f-2) m \sin^2\(\pi \frac{Q_{EM}^{tot}-Q_{EM,1}^{tot}}{(N_f-2)}\)+m\frac{\alpha}{2}\(\frac{1}{2}-\frac{\theta}{2\pi}\)^2.
\end{align}
The first two terms are the mechanical energies associated with $\phi_{b,1}$ and $\phi_{\ell,1}$, the third term is the mechanical energy associated with the rest of the $N_f-2$ fermions, and the last term is the electromagnetic energy.
$Q_{EM,1}^{tot}$ is a free parameter and, upon minimizing the energy, we see that $Q_{EM,1}^{tot}=Q_{EM}^{tot}$ so that only the first generation is excited. But if $|Q_{EM}^{tot}| > 1/2$, it is not difficult to show from Eq.~\ref{Eq: Nf total charges} that either $|Q_{B,1}^{tot}|$ or $|Q_{L,1}^{tot}|$ will be greater than 1 and so the state is classically unstable decay via fermion emission.  As a result, $\DD_{1,(1,0,...,0)}$ is the only $B-L$ charged bound state that could possibly be stable.  Therefore a decay of $\DD_{0,\vec 0}$ into this state is only possible if $0<\theta<2\pi$. If $\DD_{1,(1,0,...,0)}$ is unstable, then the $\DD_{0,\vec 0}$ state must decay directly into the $\DD_{2,\vec 0}$ state through the emission of two solitons, the energy of which can be related to the energy of the $\DD_{0,\vec 0}(\theta - 2\pi)$ state via
\begin{align}
\label{Eq: D200000 Energy}
E^{N_f}_{2,\vec 0}(\alpha,\theta)=&mN_f\sin^2\(\frac{\pi}{N_f}\(1-\frac{\theta}{2\pi}\)\)+m\frac{\alpha}{2}\(1-\frac{\theta}{2\pi}\)^2 .
\end{align}
All of this means that in order to find $\theta_c$ for the $Sp(N_f)$ theory, one must solve 
\bea
E_{0,\vec 0}(\alpha,\theta_c)= \begin{cases} E^{N_f}_{1,(1,0,...,0)}(\alpha,\theta_c) + m &\text{if} \quad 0<\theta_c<2\pi\\
 E^{N_f}_{2, \vec 0}(\alpha,\theta_c) + 2m &\text{if} \quad \theta_c \geq 2\pi .
\end{cases}
\eea
We can get the analytic expression for $E_{0,\vec 0}$ using Eq.~\ref{Eq: E0000000=E00} and Eq.~\ref{Eq: Sp2 Analytic Energy} to find
\begin{align}
\label{Eq: SpN q0 Energy}
E_{0,\vec 0}(\alpha,\theta)=&mN_f\sin^2\(\frac{\theta}{2N_f}\)+m\frac{\alpha}{2}\(\frac{\theta}{2\pi}\)^2 ,
\end{align}
which can be used to solve for $\theta_c$.  In the $N_f\gg1$ and $\alpha\ll 1$ limit, we find $\theta_c=N_f\sin^{-1}(2/\pi)\approx 0.69 N_f$. This approximation and the exact solution are compared in Fig.~\ref{Fig: thetac Nf}, showing good agreement. 

Now we turn to the question of how many stable states there are at $\theta=0$. At first this seems difficult to answer, especially in the $Sp(N_f)$ theory where there are a large number of states due to the $B-L$ charges. However, consider Eq.~\ref{Eq: D100000 Energy}, which shows that the $q=1$, $|\vec n_{B-L}|=1$ state has energy $E\geq m$. The $q=0$ state trivially has energy $0$ at $\theta=0$, so this state can never be stable. This easily extends to any state that is charged under any of the $B-L$ symmetries, and we conclude that no stable $B-L$ charged state can exist at $\theta=0$.

We have now reduced the problem to finding stable electrically-charged flavor-neutral states at $\theta=0$.  This information can be extracted from $\theta_c$ using the field redefinitions in Eqs.~\ref{Eq: SO(2) symmetry relation} and~\ref{Eq: Sp(2) symmetry relation}. By the definition of $\theta_c$ and suppressing the $\vec n_{B-L}$ index,
\bea
\label{Eq: thetac stab}
D_{0}(\theta)\quad \text{is stable if} \quad |\theta|<\theta_c .
\eea
We can take $\theta=-\pi q$ for some integer $q$ (an even integer for $Sp(N_f)$ and use Eq.~\ref{Eq: SO(2) symmetry relation} or Eq.~\ref{Eq: Sp(2) symmetry relation} to rewrite this as the statement 
\begin{align}
\label{Eq: thetac stab after sym SON}
SO(N_f):& \quad D_{q}(0)\quad \text{is stable if} \quad |q|<\frac{\theta_c}{\pi}\\
Sp(N_f):& \quad D_{q}(0)\quad \text{is stable if} \quad |q/2|<\frac{\theta_c}{2\pi} .
\end{align}
By counting the number of states satisfying these conditions, we can conclude that
\begin{align}
SO(N_f): \quad N_{st}=1+2\left \lfloor{\frac{\theta_c}{\pi}}\right \rfloor\rar\frac{3.8N_f}{\pi} \quad\text{if}\quad N_f\gg1,\; \alpha N_f/2\ll1 \\ \nn
Sp(N_f): \quad N_{st}=1+2\left \lfloor{\frac{\theta_c}{2\pi}}\right \rfloor\rar\frac{N_f}{\pi}\sin^{-1}(2/\pi)\quad\text{if}\quad N_f\gg1,\; \alpha N_f/2\ll 1 .
\end{align}

\subsection{Mass Heirarchies}
Now let us discuss the same scenarios where there is some number $n$ of light fermions with mass $m$ and some number of heavy fields $N_f-n$ with mass $M$.

Based on our intuition from Sec.~\ref{Sec: Nf2}, it is quite easy to extend the above analysis to this case for $SO(N_f)$. In the $SO(N_f)$ theories, the boundary conditions do not place any restrictions on the amount of electric charge that needs to be in any particular field (only on the sum of all the charges). As we saw in Sec.~\ref{Sec: Nf2}, the system prefers to give charge to lighter fields since it costs less energy to excite them. Thus, in the limit $M\gg m$, the system will give no charge to the heavy fermions and will instead split it all among the light fermions. This effectively leaves us with an $N_f\rar n$ theory of equal-mass fermions, and so all of the above results are valid for these hierarchical theories with the replacement $N_f\rar n$.

Extending to arbitrary numbers of heavy and light fermions for $Sp(N_f)$ is not trivial since the physics will depend on exactly how the heavy and light fermions are paired into the $N_f/2$ families. To keep the discussion relatively simple, let us only consider the case of either one heavy fermion or one light fermion. 

\paragraph{One Heavy Fermion} : 
If we have only one heavy fermion, then it will cost energy to give a charge to this fermion and the system will prefer to not excite it.  Since the system cannot excite the associated light fermion and the heavy fermion separately, this essentially removes this fermion pair from the system and so all of the above results translate with $N_f\rar N_f-2$.

\paragraph{One Light Fermion} : 
If we have only one light fermion, then the system will always prefer to give as much charge as possible to this light fermion and none to the heavy fermions. However, the boundary conditions require that if charge is given to the light fermion, then charge must also be given to its heavy partner.  On the other hand, there is no requirement from the boundary conditions to give charge to the other fermions, and so they remain unexcited. This effectively reduces the system to the $Sp(2)$ theory as long as one does not wish to consider monopoles with $B-L$ charge in the heavy families.

Amusingly, if all of the heavy fermions are not equal in mass, then $\theta$ dynamics can instead be dominated by the lightest pair of fermions, as opposed to the lightest fermion.  Namely if the light fermion's heavy partner is too heavy, it will be energetically more favorable to simply excite the other, not-quite-as-heavy fermion pairs instead.

\section{Fun with the $SU(5)$ GUT Monopole} \label{Sec: SU5}

The minimal $SU(5)$ monopole can be understood as a monopole resulting from the spontaneous breaking of an $SU(2)$ subgroup of $SU(5)$.  The bosonized picture surrounding an $SU(5)$ monopole has been derived several times~\cite{Callan:1982au,Callan:1983tm,Dawson:1983cm}, so we will simply summarize the results below.

The fundamental monopole of $SU(5)$ is a monopole under the $SU(2)$ subgroup
\bea
\vec T = \frac{1}{2} \begin{pmatrix}
           0 & & & \\
           & 0 & & \\
           & & \vec \sigma & \\
           & & & 0
         \end{pmatrix} ,
\eea
where the top-left 3x3 matrix is the color subgroup and the bottom-right 2x2 matrix is electroweak.  As expected, $SU(3) \times SU(2) \times U(1)/\mathbb{Z}_6$ are intertwined under this choice of $SU(2)$ subgroup. Under this spontaneously broken $SU(2)$, the $10$ and $\overline 5$ decompose into
\bea
\begin{pmatrix}
           e \\
           d_3^c
         \end{pmatrix} \qquad \begin{pmatrix}
           d_3 \\
           e^c
         \end{pmatrix}\qquad \begin{pmatrix}
           u_1^c \\
           u_2
         \end{pmatrix}\qquad \begin{pmatrix}
           u_2^c \\
           u_1
         \end{pmatrix} 
\eea
where the subscripts are color indices.  To proceed, we neglect off-diagonal generators and consider only the diagonal generators corresponding to
\begin{eqnarray}
\lambda_3 &=& \frac{1}{2} \, \text{diag}(1,-1,0,0,0) \\
\lambda' &=& \frac{1}{2 \sqrt{2}} \, \text{diag}(-1,-1,1,1,0) = \frac{1}{\sqrt{8}} Q_{E\&M} - \sqrt{\frac{2}{3}} \lambda_8 \\
\lambda_Z &=& \frac{1}{2 \sqrt{10}} \, \text{diag}(1,1,1,1,-4) . 
\end{eqnarray}
These generators were chosen such that they are orthogonal to the $SU(2)$ so that charges of these groups do not undergo the Witten effect.  The $Z$ boson obtains a mass and is neglected, though including it will not change anything that we discuss.  The remaining diagonal generators can be integrated out.  After bosonization, the remaining part of the gauge sector is 
\bea \label{Eq: SU(5) L}
\mathcal{L}_{\rm gauge} = - \frac{g^2}{128 \pi^3 r^2} \left [ \l \phi_{u_1} + \phi_{u_2} - \phi_{d_3} - \phi_{e} - 2 \theta_5 \r^2 + \l \phi_{u_1} - \phi_{u_2} \r^2 + \frac{1}{2} \l \phi_{e} - \phi_{d_3} \r^2 \right ] ,
\eea
where $\theta_5$ is the $SU(5)$  $\theta$ parameter.
Aside from this, there are mass terms and kinetic terms.  The boundary conditions at the origin are
\begin{eqnarray} \label{Eq: su5 boundary}
\partial_r \phi_{u_1} = -\partial_r \phi_{u_2} &\qquad& \phi_{u_1} = \phi_{u_2} \qquad \text{mod} \quad 4 \pi \\
\partial_r \phi_{d_3} = -\partial_r \phi_{e} &\qquad& \phi_{d_3} = \phi_{e} \qquad \text{mod} \quad 4 \pi . 
\end{eqnarray}
At this level, the $SU(5)$ monopole is simply an $N_f = 4$ version of the theories discussed in Sec.~\ref{Sec: Nfgeq4} with some extra gauged $U(1)$s.

In what follows, we will mostly limit ourselves to qualitative discussions of the properties of the $SU(5)$ monopole/dyon.  Any quantitative discussion is necessarily suspect because of several issues.  When deriving Eq.~\ref{Eq: SU(5) L}, all off-diagonal gauge bosons, as well as the Z boson, were all neglected.  Additionally, the $SU(5)$ monopole involves the Higgs boson and thus has additional structure (that was ignored) on scales smaller than the Higgs mass.  Finally, confinement is an extremely important low-energy feature that cannot be captured by this analysis.

\paragraph{Massless down quark solution to the Strong CP problem}: If the down quark were massless, it would solve the strong CP problem.  However, despite the fact that the down quark is massless, $\theta_5$ is still physical around monopoles.  If the down quark were massless, then $\phi_d$ could be shifted without worrying that any phase would re-appear in the mass term.  However, shifting $\phi_d$ would move $\theta_5$ to the boundary condition and the other gauge boson term so that $\theta_5$ is always physical.  The necessity of having charge 0 under the $\lambda'$ generator forces $\theta_5$ dependence even in the limit of a massless quark.  Observables such as the charge density and mass of the monopole would have $\theta_5$ dependence.  On the other hand, if both the down quark and electron (or just the up quark) were massless, then $\theta_5$ could be completely shifted away and would be unphysical.

\paragraph{Lack of bound states}: We next ask whether or not there is a quark/electron monopole bound state.  Of course, due to confinement we cannot say anything for certain.  
If we increase the Higgs vev such that $m_u = m_d = m_e > 1$ GeV, then we are in the $Sp(N_f = 4)$ situation studied before.  Given that we have experimentally found that $\theta_5 \approx 0$, all of the bound states with non-zero flavor symmetries are unstable.  The dyonic bound state with $\theta_5 = 2\pi$ is unstable due to electromagnetic effects to decaying via emission of 2 up quarks or a positron and an anti-down quark.  It is unclear if off-diagonal generators or confinement can stabilze this bound state.

\paragraph{Effects of $\theta_{\rm E\& M}$}: While the QCD $\theta_{QCD}$ angle is consistent with zero experimentally, it is possible that $\theta_{\rm E\& M}$ is non-zero and large.  A situation such as this can arise if both the $SU(5)$ $\theta_5$ angle and phases in the Standard Model masses were 0, and if we added a new vector-like multiplet in the fundamental representation.  If the mass term for the charge $1$ part ($E$) had a phase, while the mass term for the colored piece ($D$) were real, then this would induce a non-zero $\theta_{\rm E\& M}$ but keep $\theta_{\rm QCD} = 0$ after it was integrated out.

The effects of a non-zero $\theta_{\rm E\& M}$ are interesting as there is no solution unless $D$ and $E$ are explicitly included in the Lagrangian!  To see this, let us rewrite Eq.~\ref{Eq: SU(5) L} when $\theta_5 = 0$ :
\begin{eqnarray} \label{Eq: SU(5) SM}
\mathcal{L}_{\rm gauge} &=& - \frac{g^2}{32 \pi^3 r^2} \big [ \, \frac{3}{8} \l \frac{2}{3} \phi_{u_1} + \frac{2}{3} \phi_{u_2} - \frac{1}{3} \phi_{d_3} - \phi_{e} \r^2 + \l \frac{1}{2}\phi_{u_1} - \frac{1}{2}\phi_{u_2} \r^2 \nn \\
&+& \l \frac{1}{2 \sqrt{3}} \phi_{u_1} + \frac{1}{2 \sqrt{3}} \phi_{u_2} - \frac{1}{\sqrt{3}} \phi_{d_3} \r^2 \big ] ,
\end{eqnarray}
where the first term comes from E\&M, and the remaining terms come from $\lambda_{3,8}$ of QCD.
Let us discuss things in terms of the color singlet proton-like state ($\phi_{u_1} = \phi_{u_2} = \phi_{d_3} \equiv \phi_p$) and the electron.  This sub-system has nice properties such as the boundary conditions $\partial_r \phi_p = \partial_r \phi_e = 0$ and $\phi_{p} = \phi_{e}$ mod $4 \pi$, as well as leaving only the electromagnetic piece of Eq.~\ref{Eq: SU(5) SM} non-zero :
\bea \label{Eq: low E}
\mathcal{L} = - \frac{3}{256 \pi^3 r^2} \l \phi_{p} - \phi_{e} + \theta_{\rm E\& M} \r^2 ,
\eea
where we have now included the $\theta$ term for E\&M.
Finite energy forces $\phi_{p} = \phi_{e} - \theta_{\rm E\& M}$ at small radii, while the boundary conditions force $\phi_{p} = \phi_{e}$.  Clearly, finite energy and the boundary conditions are inconsistent.

The resolution to this apparent paradox is to examine what happens at the scale where $E$ is integrated out.  Including this new particle changes the Lagrangian to
\bea 
\mathcal{L} = - \frac{3}{256 \pi^3 r^2} \l \phi_{p} - \phi_{e} - \phi_E \r^2 + m_E^2 \cos \l \phi_E + \theta_{\rm E\& M} \r . 
\eea
At the scale $r \sim 1/m_E$, $E$ deposits a charge $\theta_{\rm E\& M}$ that is subsequently screened by the proton and electron.  In this manner and in the $m_E \gg m_p, m_e$ limit, the boundary condition effectively changes from $\phi_{p} = \phi_{e}$ to $\phi_{p} = \phi_{e} - \theta_{\rm E\& M}$.  In practice, the screening of the electrons and protons is so inefficient $(m_{p,e}/m_E)^{\alpha/4 \pi} - 1 \approx 0$ that it may as well not occur.  All of this results in the surprising fact that the dependence of the monopole mass and other properties on $\theta_{\rm E\& M}$ will be dominantly be driven by $m_E$, namely undiscovered UV physics, as opposed to any of the observed lighter particles.

\paragraph{Velocity dependence of the Callan Rubakov effect}: Of great phenomenological interest is the rate at which baryons turn into electrons.  When thrown in at low energies, chirality violation due to the mass term is the dominant effect and $B-L$ violation is suppressed.  At high energies, the boundary conditions are the dominant effect and $B-L$ violation occurs.  There is a critical velocity at which $B-L$ violation goes from exponentially unlikely to saturating unitarity.  We now briefly estimate this velocity (see Ref.~\cite{Dawson:1983cm} for related work).

In the low energy limit, the monopole is described by protons and electrons with the electromagnetic term shown in Eq.~\ref{Eq: low E} and the boundary conditions $\partial_r \phi_p = \partial_r \phi_e = 0$ and $\phi_p = \phi_e$.  We will take $\theta_{\rm E\& M} = 0$ in order to simplify the discussion.   We have numerically solved the e.o.m. with Eq.~\ref{Eq: low E} by sending in solitons corresponding to protons and determined the critical velocity at which the outgoing particle is a positron.  While we were not able to numerically solve the problem for realistic values of $\alpha = 1/137$, we were able to find the scaling
\bea \label{Eq: vc}
v_c \approx 0.2 \sqrt{\alpha} .
\eea
To obtain this numerical result, we took the electron to be massless and varied $\alpha$ between 0.1 and 5.  We fit the scaling of $v_c$ and obtained Eq.~\ref{Eq: vc}.  Extrapolated to $\alpha = 1/137$, we find that 
\bea \label{Eq: vcSM}
v_c \approx 2 \times 10^{-2}.
\eea
For velocities faster than this value, we expect proton conversion to electrons by $SU(5)$ monopoles to be unsuppressed, while we expect conversion to be exponentially suppressed for velocities smaller than this.

We can make a rough estimate to understand the $\alpha$ scaling of the critical velocity.  Let us imagine that a baryon is incoming on the monopole with velocity $v$.  Eventually, at a critical radius $R_c$, the electromagnetic energy of a spherical charged shell is equal to its original kinetic energy ($\alpha/(8 \pi R_c) \sim m_p v^2/2$) and the soliton bounces back.  If the electron and proton are both effectively massless at this critical radius ($R_c <  1/m_p < 1/m_e$), then the reflection of the baryon is similar to the massless limit, and $B-L$ is violated.  Otherwise, chirality violation from the mass term is more important, and no $B-L$ violation occurs.  Solving for the critical velocity when this turnover occurs, we find $v_c \sim \sqrt{\alpha/(4 \pi)}$, which is closer to the numerical value, Eq.~\ref{Eq: vc}, than it has any right to be.

Finally, we note that Eq.~\ref{Eq: vcSM} should be taken with a grain of salt.  As emphasized in the previous paragraph, conversion finally occurs when $R_c \sim 1/m_p$.  Unfortunately, confinement becomes important at distances $\sim 1/m_p$, and our EFT treatment of the proton breaks down.  We expect that a proper treatment would only change Eq.~\ref{Eq: vcSM} at the $\mathcal{O}(1)$ level.  The lack of conversion at smaller velocities still lies in the realm of validity of our EFT approach.  For another approach towards modeling the proton electron system, see Ref.~\cite{Callan:1983nx}.

\section{The Massless Limit} \label{Sec: semiton}

Monopoles with massless fermions have been a source of confusion for many years.  One of the original confusions came from the Witten effect discontinuously vanishing when a fermion mass was non-zero versus zero.  This confusion was due to an order of limits issue and was resolved by starting with a massive fermion and sitting at a fixed distance from the monopole.  The charge implied by the Witten effect was distributed within a radius of order $1/m$.  As the fermion mass was smoothly taken to zero, this charge density would sweep over the fixed-distance observer, and the effective charge of the monopole would smoothly transition to zero~\cite{Rubakov:1981rg,Rubakov:1982fp,Rubakov:1983sy,Callan:1982ah,Callan:1982au,Wilczek:1981dr}. 

More recently, another confusing aspect of monopoles with massless fermions has resurfaced.  This is the issue of semiton states.  The crux of the issue is that if we imagine a spontaneously broken $SU(2)$ theory with $N_f  \ge 4$ massless fermions and throw an s-wave fermion at the monopole, there is no outgoing s-wave state that has the correct quantum numbers.  Matching the quantum numbers require the existence of a state, called the semiton, that appears to have the quantum numbers of $2/N_f$ of each of the fermions.  One of the many confusing aspects of the appearance of the semiton is that there appears to be a new asymptotic state that does not have the quantum numbers of any of the fermions in the system.  Far from the monopole, we would think that we could quantize the theory in the standard manner and see only the flavorful fermionic excitations.  The puzzle about the origin of this semiton state and whether it makes any sense has prompted many papers and proposed resolutions~\cite{Polchinski:1984uw,Callan:1983tm,Kitano:2021pwt,Csaki:2020inw,Csaki:2021ozp,Brennan:2021ewu,Hamada:2022eiv,vanBeest:2023dbu,Brennan:2023tae,vanBeest:2023mbs,Csaki:2024ajo,Loladze:2024ayk}.  

In this section, we seek to answer a different question than previous works.  Since the semiton states exist in the massless limit and do not exist in the massive limit, we seek to answer the question where do semitons come from?  
We approach the problem in much the same way as the Witten effect was originally studied. Namely, we first take the massive theory with equal masses and sit a fixed distance away from the monopole.  We then take the mass to be small and see what the fixed distance observer sees.

We begin with $N_f \ge 4$ and mass terms that preserve the $SO(N_f)$ flavor symmetry.  The stars of the show are the bound states without fixed flavor quantum numbers that were found in Sec.~\ref{Sec: Nfgeq4}.  These bound states extend over a distance $1/m$ from the monopole.  Initially, the asymptotic observer is sitting at a distance $r \gg 1/m$ from the monopole, and the only states near the observer are the quantized flavorful fermion states.  Now the masses of the fermions are taken to become smaller and smaller.  Eventually, the bound state reaches the observer and new ``aymptotic" states are now present in the theory: the flavorless bound state excitations!  This provides some physical understanding as to the origin of the semiton states and how they appear as genuine ``asymptotic" states in the massless limit.

\section{Conclusion}
\label{Sec: conclusion}

In this paper, we studied (multi) fermion-monopole bound states.   Depending on the  UV flavor symmetry, a whole tower of dyonic bound states can be stable.  In the large $N_f$ limit, most of these bound states are the fractional $\theta = 2 \pi \mathbb{Z}/N_f$ fermion vacuum solutions, familiar from the Witten effect, distributed among the $N_f$ fermions.  As a result, these bound states typically carry no flavor quantum number.  Amusingly enough, these bound states represent new final state possibilities for Callan-Rubakov scattering processes.  In an $SO$ flavor symmetric theory, a low velocity fermion incident on a monopole can be numerically solved in its bosonized form, and the final state is a bound state plus many soft photons, in which no fermion is emitted.

We also studied the charge distribution and stability of these bound states. The charge distribution is not exponentially localized around the magnetically charged object, but only polynomially localized.  Once the flavor symmetry is broken by allowing for different masses, phase transitions occur as states became stable or unstable.  We found that typically once bound states were quantum mechanically unstable, they would quickly become classically  unstable as $\theta$ was varied.  We hope that one day these dynamics will be relevant for the real world.

\acknowledgments{
AH and CR were supported in part by the NSF grants PHY-1914480 and by the Maryland Center for Fundamental Physics (MCFP).  We that the anonymous referees for their helpful comments and suggestions.
}

\appendix

\section{Bosonization Around Monopoles} \label{Sec: Bosonization Details}

\subsection{Fermions around a Monopole}
\label{Sec: 2D reduce}
To compute the dyonic bound states of fermions around a monopole, we will need a framework in which to study and compute the fermion fields. In this section, we will describe the process of computing fermion fields in a 't Hooft-Polyakov monopole background. We follow the a standard procedure~\cite{Callan:1982au,Callan:1982ah} by first reducing the full 1+3D theory of fermions to a 1+1D theory of fermions by considering only the total angular momentum $\vec J=0$ mode of the fermions which interacts with the monopoles core. Then we make use of a duality between fermions and bosons in 1+1D theories to rewrite our theory of $N_f$ fermions as a theory of $N_f$ scalars through a process called bosonization. These scalar fields were then used to compute the fermion states in the main text.

\subsubsection{Gauge Sector and the Monopole Background}
The 't Hooft-Polyakov background is described by the following background gauge field, $A_\mu^a$, and scalar field, $\Phi^a$.
\bea
\label{Eq: Monopole Background}
A^a_0 = \dot\lambda(t,r)\hat r_a \quad A_i^a=- \epsilon_{i a b} \hat r_b \frac{A(r)}{er} \quad \Phi_a= v\hat r^a Q(r)
\eea
where $a,b$ are gauge indices, $i$ is a spatial index, $\hat r$ is a unit radial vector, $e$ is the gauge coupling, and $A(r)=Q(r)=1$ for $r \gg r_M$ and $A(r)=Q(r)=0$ for $r \ll r_M$, where $r_M$ is the radius of the monopole core.  We will write the generators of the $SU(2)$ gauge group as Pauli spin matrices over 2, $\tau_a/2$. 
One can show that this field yields electric and magnetic fields far from the monopole
\bea 
B_{a,i}=\frac{\hat r_a\hat r_i}{er^2} \quad E_{a,i}= - \hat r_a\hat r_i \dot\lambda'(t,r)
\eea
Where it is understood that the primes are derivatives with respect to $r$ and the dots are derivatives with respect to time. $\dot\lambda'(r,t)$ describes a radial electric field degree of freedom around the monopole. Let us plug this background gauge field into the pure gauge sector of the Lagrangian to give 
\bea
\label{Eq: Gauge Lagrangian}
L_{Gauge}=\int_0^\infty dr (4\pi r^2)\( \frac{1}{2}(E_{a,i}^2-B_{a,i}^2)-\frac{\theta e^2}{8\pi^2}E_{a,i}B_{a,i}\) .
\eea
 We will only keep the terms that contain $\lambda$ as it is the only remaining degree of freedom in the gauge sector. 
\bea
\label{Eq: Gauge Lagrangian 2D}
L_{Gauge}=\int_0^\infty dr 2\pi r^2 \dot\lambda'^2+\frac{\theta e}{2\pi}\dot\lambda'
\eea
\subsubsection{$J=0$ Fermion Modes}
This background gauge field is rotationally symmetric as long as spatial rotations about some axis by an angle $\theta$ must be compensated by a gauge rotation by $-\theta$ about that same axis in gauge space. This breaks rotational symmetry and gauge symmetry into its diagonal subgroup (crossed with the electromagnetic $U(1)$ group) where one transforms both the gauge and rotational degrees of freedom equally and opposite. This implies that the unitary operator $U(\theta_a)=\exp(-i J^a\theta^a)$, defined as 
\bea
e^{-i\vec J\cdot \vec \theta}=e^{-i(\vec L+\vec S)\cdot \theta}e^{i\frac{\vec \tau}{2}\cdot (-\theta)}=e^{-i(\vec L+\vec \sigma/2+\vec \tau/2)\cdot \theta}
\eea
where $\vec \tau$ and $\vec \sigma$ are both Pauli matrices acting on gauge and spin indices respectively, leaves the Lagrangian invariant. We can therefore interpret $\vec J$ as the total angular momentum operator on fermions. We are interested in fermions states that interact with the monopole core and thus have $\vec J=0$. One can find the most general form of the $\vec J=0$ state is a superposition of two states
\bea
\label{Eq: J=0 Fermions}
\psi_{J=0}=g(r,t)\psi_{J=0;L=0}+p(r,t)\psi_{J=0;L=1}=\frac{g(r,t)+p(r,t)(\hat {\vec r}\cdot \vec \sigma)}{\sqrt{4\pi r^2}}|\vec \tau+\vec \sigma=0\rangle
\eea
where $|\vec \tau+\vec \sigma=0\rangle$ is the spin-isospin singlet state and can be thought of as a $2\times 2$ matrix $i\sigma^2_{\alpha \alpha_2}/\sqrt{2}$ with doublet gauge index $\alpha_2$ and Lorentz spin index $\alpha$.
In App.~\ref{Appendix: Fermions and Monopoles} we derive the following identities for any $\psi$ in the $\vec J=0$ mode. 
\begin{align}
\label{Eq: 2D kinetic}
i\psi^\dag \bar\sigma^\mu\partial_\mu\psi=&\frac{i}{4\pi r^2}\(\bar \xi \bar\gamma^\mu\partial_\mu\xi - \frac{1}{r}\bar \xi \bar\gamma^5\xi\)\\
\label{Eq: Current 2D reduction}
j^0\equiv \psi^\dag \bar\sigma^0 \psi=&\frac{1}{4\pi r^2}\bar\xi\bar\gamma^0\xi\\ \nn
j^r\equiv \hat r_i\psi^\dag \bar\sigma^i \psi=&\frac{1}{4\pi r^2}\bar\xi\bar\gamma^1\xi\\
\label{Eq: EM Current 2D reduction}
j_{EM}^0\equiv \psi^\dag (\hat {\vec r}\cdot\vec \tau)\bar\sigma^0 \psi=&\frac{1}{4\pi r^2}\bar\xi\bar\gamma^1\xi\\ \nn
j_{EM}^r\equiv \hat r_i\psi^\dag (\hat {\vec r}\cdot\vec \tau) \bar\sigma^i \psi=&\frac{1}{4\pi r^2}\bar\xi\bar\gamma^0\xi\\
\label{Eq: Monopole Term}
e\psi^\dag \bar\sigma^i\frac{\tau^a}{2}\psi A_i^a=&\frac{i}{4\pi r^2}\frac{A(r)}{r}\bar \xi\bar\gamma^5\xi 
\end{align}
where $\xi$ and $\bar \gamma^\mu$ are 1+1D fermions and 1+1D gamma matrices respectively defined as 
\bea 
\label{Eq: 2D Spinor and gammas}
\xi=\frac{1}{\sqrt{2}}\mat g-p\\-i(g+p)\rix\qquad \bar\gamma^0=\sigma_2\quad \bar\gamma^1=i\sigma_1\quad \bar\gamma^5=\sigma_3 \, .
 \eea
This reduces the fermion sector of the 1+3D Lagrangian to a 1+1D Lagrangian
\bea
\label{Eq: 2D Fermion Lagrangian 1}
L_{\rm Fermion}=\int d^3\vec r\; i\psi_i^\dag \bar \sigma^\mu D_\mu\psi_i=\int_0^\infty dr\; i\bar\xi_i\bar\gamma^\mu\partial_\mu\xi_i+\frac{e}{2}\dot\lambda(r,t)\bar\xi_i\bar\gamma^1\bar\xi_i+i\frac{A(r)-1}{r}\bar \xi_i\bar\gamma^5\xi_i
\eea
In App.~\ref{Appendix: Boundary} we show that if the solution is to be well-behaved and finite that the last term in this Lagrangian imposes the boundary condition $p(r=r_M)=0$, where $r_M$ is the mass of the monopole. Physically, this reflects $p$ corresponding to the $\vec L=1$ portion of the $\vec J=0$ mode and so is pushed away from the core once rotational symmetry and gauge symmetry become decoupled at $r=r_M$. With this boundary condition established, since we are interested in the fermion state outside of the core, we can set $A(r)=1$ and remove this term from the Lagrangian. 

Next, we insert the $\vec J=0$ modes into the two types of mass terms. Our two mass terms are
\bea
\delta\LL_{SO(N_f)} = - \frac{y}{2} \psi \Phi \psi \qquad \delta\LL_{Sp(N_f)}= - m \psi_A \psi_B - y \psi_A \Phi \psi_B 
\eea
for fermions $\psi_A$ and $\psi_B$. The flavor index $A,B$ is suppressed in the $SO$ mass term. The masses were chosen such that the fermion determinant is positive and hence the only source of CP violation is $\theta$.  For a quick reminder why the fermion determinant is what is relevant for CP violation, see Refs.~\cite{Vafa:1983tf,Vafa:1984xg}.
We can insert Eq.~\ref{Eq: J=0 Fermions} into these equations to find
\begin{eqnarray}
\label{Eq: SON Massterm 2D pt 1}
4\pi r^2\delta\LL_{SO(N_f)}&=&-m\frac{(g_i+p_i)(g_i-p_i)}{2}\\
4\pi r^2 \delta\LL_{Sp(N_f)}&=&+(m - y v) \(\frac{(g_{B,i}+p_{B,i})(g_{A,i}-p_{A,i})}{2} \)  \\
& &- (m + y v ) \( \frac{(g_{A,i}+p_{A,i})(g_{B,i}-p_{B,i})}{2}\) \nn
\label{Eq: SpN Massterm 2D pt 1}
\end{eqnarray}
Neither of these looks much like a mass term for our 1+1D fermions. To rectify this, we can redefine our $\xi$'s in both theories so that the above expressions better map onto mass terms for our 1+1D fermions. Noting that in our basis of 1+1D gamma matrices, 
\bea
\label{Eq: 2D Mass Term}
L_{mass}=-\int_0^\infty dr m_i \bar \xi_i\xi_i
=-m_i \int_0^\infty dr \(i\xi_-^*\xi_++h.c.\)
\where \xi=\mat \xi_+\\\xi_-\rix,
\eea
we can see that if we redefine $\xi$ for the $SO(N_f)$ type mass terms as
\bea
\label{Eq: SON xis bl}
\xi_i=\frac{1}{\sqrt{2}}\mat g_i-p_i\\ g_i^*+p_i^*\rix 
\eea
and define $\xi_{\ell,i}$ and $\xi_{b,i}$ for the $Sp(N_f)$ type mass terms as
\bea
\label{Eq: SpN xis bl}
\xi_{b,i}=\frac{1}{\sqrt{2}}\mat g_{A,i}-p_{A,i}\\ g_{B,i}^*+p_{B,i}^*\rix \qquad \xi_{\ell,i}=\frac{1}{\sqrt{2}}\mat g_{B,i}-p_{B,i}\\ g_{A,i}^*+p_{A,i}^*\rix 
\eea
the mass terms in Eqs.~\ref{Eq: SON Massterm 2D pt 1} and~\ref{Eq: SpN Massterm 2D pt 1} take the form $m\bar\xi\xi$ for the various 1+1D fermions.  

Because it will become relevant in a moment, let us make a detour and quickly discuss an often neglected part of the discussion of the physically observable angle $\bar \theta$.  As is often discussed, $\bar \theta$ depends on the choice of $\theta$ and the phase of mass terms.  What is often not discussed is that $\bar \theta$ also depends on the choice of measure, as it is what determines what a positive mass with no phase is equal to.  Conventionally, one chooses the measure for Dirac fermions $D \bar \Psi D \Psi$ so that a positive mass with no phase is $\delta L = - m \bar \Psi \Psi$~\cite{Vafa:1983tf,Vafa:1984xg}.  However if one chooses a measure $D \Psi D \bar \Psi$, then a positive mass corresponds to $\delta L = m \bar \Psi \Psi$.  An amusing tidbit is that for the theory we consider, $SU(2)$ with two fundamentals, the conventional choice of measure $D \bar \Psi D \Psi$ isn't unique as there is a relative minus sign depending on how one constructs the Dirac fermion out of our Weyl fermions.

And now, back to our regularly scheduled programming.  An astute reader, will notice that taking Eqs.~\ref{Eq: SON xis bl} and~\ref{Eq: SpN xis bl} and plugging them directly into Eq.~\ref{Eq: 2D Mass Term} only reproduces  Eqs.~\ref{Eq: SON Massterm 2D pt 1} and~\ref{Eq: SpN Massterm 2D pt 1} up to factors of $i$ and $-1$.   The extra step required to match the two theories is a field redefinition and an anti-commutation of zero modes so that the measures of the two theories agree.  When writing our 2D theory, the traditional choice of measure is $D \bar \xi D \xi$ which depends on the choice of $\bar \gamma^0$.  The 4D theory has $\sigma^0 = \identity$ while the 2D theory has $\bar \gamma^0 = \sigma_2$, giving the missing factors of $i$.

The $-1$ mismatch can be seen in the purely fermionic theory as it comes from the $\epsilon$ tensor needed to contract two fundamentals of $SU(2)$.  The reason why one of the two fermion pairs appears to have a ``wrong" sign mass is that the fermion measure involves integrating the zero modes of this pair of fermions in the opposite order than the other pair (as can be seen if one chose to write one of the two fermion pairs as an anti-fundamental as opposed to a fundamental).  This opposite ordering introduces a minus sign when calculating the fermion determinant.  These two minus signs, one in the measure and one in the mass, cancel and the resulting determinant is positive.  When one transitions to the 2D theory, enforcing the traditional ordering of $D \bar \xi D \xi$ gives the missing minus sign.

These redefinitions of the $\xi$'s will naturally cause changes to the other terms in the Lagrangian in Eq.~\ref{Eq: 2D Fermion Lagrangian 1}. However, it is not difficult to show that the kinetic terms are unchanged and the currents change as follows
\bea
\label{Eq: SON currents}
\text{$SO(N_f)$:} \quad \bar\xi_i \bar \gamma^0\xi_i\rar \bar\xi_i \bar \gamma^1\xi_i \quad \bar\xi_i \bar \gamma^1\xi_i\rar \bar\xi_i \bar \gamma^0\xi_i
\eea
\begin{align}
\label{Eq: SON currents}
\text{$Sp(N_f)$:} \quad \bar\xi_{A,i}\bar\gamma^0\xi_{A,i}+\bar\xi_{B,i}\bar\gamma^0\xi_{B,i}\rar& \bar\xi_{b,i}\bar\gamma^1\xi_{b,i}+\bar\xi_{\ell,i}\bar\gamma^1\xi_{\ell,i} \\ \nn \bar\xi_{A,i}\bar\gamma^1\xi_{A,i}+\bar\xi_{B,i}\bar\gamma^1\xi_{B,i}\rar& \bar\xi_{b,i}\bar\gamma^0\xi_{b,i}+\bar\xi_{\ell,i}\bar\gamma^0\xi_{\ell,i} \\
\nn
\bar\xi_{A,i}\bar\gamma^0\xi_{A,i}-\bar\xi_{B,i}\bar\gamma^0\xi_{B,i}\rar& \bar\xi_{b,i}\bar\gamma^0\xi_{b,i}-\bar\xi_{\ell,i}\bar\gamma^0\xi_{\ell,i} \\ \nn\bar\xi_{A,i}\bar\gamma^1\xi_{A,i}-\bar\xi_{B,i}\bar\gamma^1\xi_{B,i}\rar& \bar\xi_{b,i}\bar\gamma^1\xi_{b,i}-\bar\xi_{\ell,i}\bar\gamma^1\xi_{\ell,i} 
\end{align}
This change in the $\xi$'s also affects the boundary conditions at the monopole core. Directly from Eq.~\ref{Eq: SON xis bl} and~\ref{Eq: SpN xis bl}, we can see, the condition $p_i(r=0)=0$ implies that at $r=0$
\begin{align}
\label{Eq: xi BC}
\text{$SO(N_f)$:}& \quad \xi_{i,+}=\xi_{i,-}^*\\
\text{$Sp(N_f)$:}& \quad \xi_{\ell,i,+}=\xi_{b,i,-}^*\quad \xi_{b,i,+}=\xi_{\ell,i,-}^*
\end{align}
where $\xi_+$ and $\xi_-$ are the upper and lower components of the spinors respectively. These conditions also imply boundary conditions on the fermion currents at $r=0$,
\begin{align}
\label{Eq: current BC}
\text{$SO(N_f)$ Theory:}& \quad \bar\xi_{i}\bar\gamma^0\xi_i=0\\
\text{$Sp(N_f)$ Theory:}& \quad \bar\xi_{b,i}\bar\gamma^0\xi_{b,i}+\bar\xi_{\ell,i}\bar\gamma^0\xi_{\ell,i}=0 \quad \bar\xi_{b,i}\bar\gamma^1\xi_{b,i}-\bar\xi_{\ell,i}\bar\gamma^1\xi_{\ell,i}=0 \, .
\end{align}
The difference in boundary conditions between the two theories is a reflection of the difference in UV flavor symmetries. In fact, other than these boundary conditions, the two theories are identical and have Lagrangian
\begin{eqnarray}
\label{Eq: 2D Fermion Lagrangian}
L &=&\int_0^\infty dr\; \bar \xi_{\ell,i}(i\slashed\partial -m_{\ell,i})\xi_{\ell,i}+\bar \xi_{b,i}(i\slashed\partial-m_{b,i}) \xi_{b,i}+\frac{e}{2}\dot\lambda(\bar \xi_{b,i}\gamma^0\xi_{b,i}+\bar \xi_{\ell,i}\gamma^0\xi_{\ell,i}) \\
\nn & & +\frac{\theta e}{2\pi}\dot\lambda'+2\pi r^2 \dot\lambda'^2
\end{eqnarray}
We have now included the possibility that the different fermions have different masses.

\subsection{Bosonization}
\label{Sec: Bosonization}
It has long been established that there is a duality between theories of bosons and fermions in 1+1D~\cite{Coleman:1974bu, Mandelstam:1975hb}. This equivalence was first realized by Colman and Mandelstam and has become a staple of field theory and condensed matter theory since. Callan used this equivalence to derive a bosonic theory for fermions around a monopole for the $Sp(N_f)$ theories with $N_f=2$~\cite{Callan:1982au,Callan:1982ah}. In this section, we generalize this process to include $SO(N_f)$ theories. Appendix~\ref{Appendix: Bosonization} provides a detailed derivation and proof of the bosonization. There we show that both theories of $N_f$ fermions, $\xi_i$, are equivalent to a theory of $N_f$ scalars $\phi_i$ in 1+1D that obey a Lagrangian 
\bea
\label{Eq: Bosonized Lagrangian}
L=\int_0^\infty dr\; \frac{1}{2}\partial_\mu\phi_i\partial^\mu\phi_i-m_i\frac{\mu_ie^{\gamma-F(r)}}{2\pi}(1-:\cos(2\sqrt{\pi}\phi_i(r,t)):)+\frac{e\dot \lambda}{2\sqrt{\pi}}\partial_r\phi_i(r,t)+\frac{e\theta}{2\pi}\dot \lambda'+2\pi r^2 \dot\lambda'^2
\eea
where there is an implied sum over all flavors $i$, the colons indicate the normal ordering of the scalar operators, $\mu_i$ is some yet-to-be-determined scale factor, all phases have been shifted into $\theta$, and masses are positive so that charge fractionalization does not occur~\cite{Jackiw:1975fn}. For the $Sp(N_f)$ theories $F(r)=0$ while for the $SO(N_f)$ theories, $F(r)=K_0(2\mu_ir)$ where $K_0$ is the zeroth modified Bessel function of the second kind. We also establish a correspondence between the fermion currents and derivatives of the scalar fields 
\bea
\label{Eq: Current Correspondence}
j_i^\mu=\bar\xi_i\gamma^\mu\xi_i=\epsilon^{\mu\nu}\frac{\partial_\nu\phi_i}{\sqrt{\pi}}
\eea
Where $\epsilon^{01}=1$ and is the antisymmetric tensor in two dimensions. This equivalence is achieved by making the substitution 
\bea
\label{Eq: Bosonized xi}
\xi_i(r,t)=Z^{1/2}(r):\mat e^{i\sqrt{\pi}\(\phi(r,t)-\int_0^r dx\dot\phi(x,t)\)}\\ e^{i\alpha}e^{-i\sqrt{\pi}\(\phi(r,t)+\int_0^r dx\dot\phi(x,t)\)}\rix:
\eea
$Z(r)$ is an overall normalization factor that is unimportant outside of the computations in appendix~\ref{Appendix: Bosonization} and $\alpha$ is some overall phase. $\alpha$, as well as the boundary conditions on the scalar fields, can be determined by applying Eqs.~\ref{Eq: xi BC} and~\ref{Eq: current BC} to Eqs.~\ref{Eq: Current Correspondence} and~\ref{Eq: Bosonized xi}.   Because we chose real positive masses for the fermions, $\alpha=0$, however if the masses had phases, $\alpha$ would be non-zero. For the $SO(N_f)$ theories the scalar fields obey the boundary condition 
\bea
\label{Eq: SON scalar BC}
\partial_r\phi_i(r=0)=0 \, .
\eea
For the $Sp(N_f)$ theories we are left with the boundary conditions 
\bea
\label{Eq: SpN scalar BC}
\phi_{\ell,i}(r=0)-\phi_{b,i}(r=0)=0 \; \text{mod}\; 2\sqrt{\pi} \quad \partial_r\phi_{\ell,i}(r=0)+\partial_r\phi_{b,i}(r=0)=0
\eea
Before using these scalar theories to compute the bound states, it will be necessary and useful to make the following three simplifications to the Lagrangian, 
\begin{enumerate}
 \item We rescale our field $\phi\rar \phi/2\sqrt{\pi}$ which makes the theory much simpler to analyze by removing the various $\sqrt{\pi}$ factors from the bosonization process
 \item We integrate out the electric field by solving the equations of motion for the electric field $\dot\lambda'$. This is done easily by integrating the third to last term in Eq.~\ref{Eq: Bosonized Lagrangian} by parts to find
 \bea
 \label{Eq: Electric Field intout}
\dot\lambda'=\frac{e}{4\pi r^2}\(\sum_i\frac{\phi_i}{4\pi}-\frac{\theta}{2\pi}\) .
\eea
Note that this has a very physical interpretation that the electric charge enclosed in a radius $r$, $Q_{EM}(r)$, is given by the term in parenthesis in Eq.~\ref{Eq: Electric Field intout}.
\item Finally, we notice that far from the monopole core, the theory reduces to a set of decoupled scalars satisfying the sine-Gordon equations. The sine-Gordon equation is known to have soliton solutions and it is these solutions that correspond to fermion states as was proven by Mandelstam~\cite{Mandelstam:1975hb}. We fix the scales $\mu_i$ by demanding that the solitons far from the monopole have energy $m_i$. More will be said about the soliton solution in App.~\ref{App: solitons}. 
\end{enumerate}
After all of these simplifications, the Lagrangian for the theories becomes
\begin{eqnarray}
\label{Eq: Bosonized Lagrangian}
L&=&\frac{1}{4\pi}\int_0^\infty dr\; \sum_i\(\frac{1}{2}\partial_\mu\phi_i\partial^\mu\phi_i-\(\frac{\pi m_i(r)}{2}\)^2(1-:\cos(\phi_i(r,t)):)\) \\ 
& &-\frac{\alpha}{8\pi r^2}\(\sum_i\phi_i-2\theta\)^2 \nn
\end{eqnarray}
where $\alpha\equiv e^2/4\pi$ and the sums over $i$ include all flavors of scalar with the conditions and definitions
\begin{align}
\label{Eq: SpN scalar BC and mass}
\text{$Sp(N_f)$ Theories:}\quad &\phi_{\ell,i}(r=0)-\phi_{b,i}(r=0)=0 \quad \text{Mod}\; 4\pi \\ \nn &\partial_r\phi_{\ell,i}(r=0)+\partial_r\phi_{b,i}(r=0)=0 \quad m_i(r)=m_i\\ 
\label{Eq: SON scalar BC and mass}
\text{$SO(N_f)$ Theories:}\quad &\partial_r\phi_i(r=0)=0 \quad m_i(r)=m_i\exp\(-K_0\(\pi^2e^{-\gamma}m_i r/4 \)/2\)
\end{align}
Note that $m_i(r)$ approaches $m_i$ for $m_ir\gg1$ and $\pi m_i\sqrt{m_i r/8} $ for $m_ir\ll 1$ in the $SO(N_f)$ theories.

\subsubsection{Solitons}\label{App: solitons}
\begin{figure}[t]
 \centering
 \includegraphics[width=0.6\textwidth]{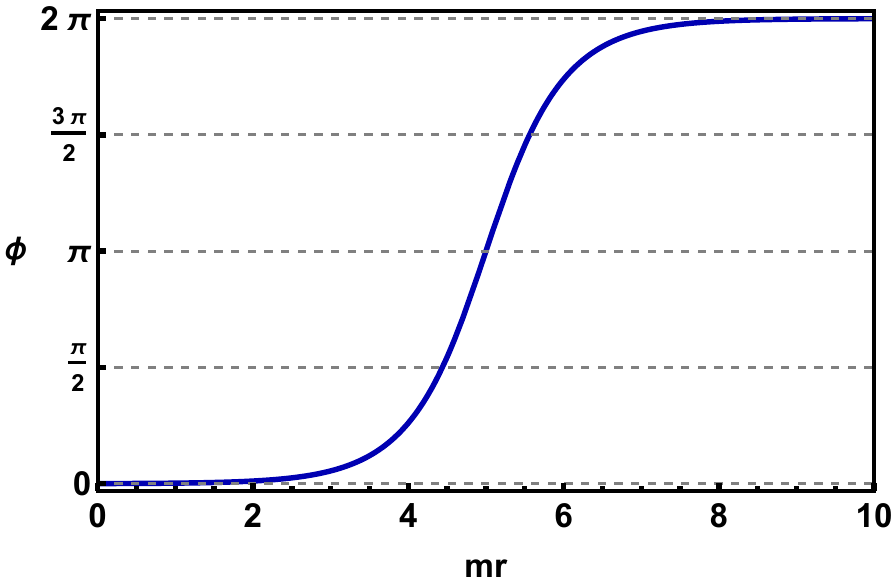}
 \caption{A plot of the soliton solution to the equations of motion in the decoupled limit. This soliton solution in the bosonized theory corresponds to a fermion in the original theory. }
 \label{Fig: Soliton}
\end{figure}

Far from the monopole, the electromagnetic term is negligible and the equations of motion become a set of decoupled sine-Gordon equations. The sine-Gordon equation famously admits solutions which run from $0$ to $2\pi$ (or $2\pi$ to 0) called solitons(or antisolitons) which have the analytic form
\bea
\label{Eq: Soliton}
\phi_{\rm soliton}(r,t)=4\arctan\(\exp\(\pi m\gamma(r-vt)/2\)\) ,
\eea
where $\gamma=(1-v^2)^{-1/2}$ is the Lorentz boost factor. A plot of this soliton solution is shown in Fig.~\ref{Fig: Soliton}. Solitons moving with velocity $v$ have energy $\gamma m$ with a soliton at rest having energy $m$. These solitons, as hinted earlier, correspond to fermions in the original theory. 

Solitons will be important in our analysis of the stability of dyonic states. These dyonic states can transition between each other via the emission of solitons so the question of stability will come down to whether or not, the difference in energy between these states is less than the energy of a soliton.

\subsubsection{Currents in Bosonized Theories} \label{Appendix: currents}

From the current correspondence in Eq.~\ref{Eq: Current Correspondence} and the relations in Eq.~\ref{Eq: Current 2D reduction}-\ref{Eq: EM Current 2D reduction}, we know that derivatives of the $\phi$'s are related to currents of the original fermions. Using these equations, one can find a direct relation between the current of the IR currents and derivatives of the bosonized scalar fields 
\bea
\label{Eq: IR currents bosonized}
4\pi r^2 \bar j_i^0=\psi_i^\dagger \overline \sigma^\mu\psi_i=\epsilon^{\mu\nu}\frac{\partial_\nu\phi_i}{2\pi}
\eea
From which one can easily compute the total charge enclosed in a sphere of radius $r$, $Q_i(r)$ for any one of these currents as: 
\bea
\label{Eq: Charge Enclosed}
Q_i(t,r)=\int_0^r dr'j^0_i(t,r')=\frac{\phi_i(r,t)-\phi_i(0,t)}{2\pi}
\eea
In the main text we find it useful to describe solutions to the theory by expressing them in terms of $Q_i(r)$ rather than the fields $\phi_i(r)$ themselves, since $Q_i(r)$ is more directly physically meaningful. We adopt the notation $Q_i^{tot}\equiv Q_i(\infty)$ to describe the total charge in a particular field. 

The currents, $j_i^\mu$ satisfy local conservation laws reflecting the fact that they correspond to, at the very least, accidental symmetries in the IR. However, many of these currents are not conserved in the UV theory. This fact can be realized by considering the boundary conditions at $r=0$. For example, consider the electromagnetic current, $j_{EM}^\mu=\frac{1}{2}\sum_ij_i^\mu$, where the $1/2$ reflects the fact that each fermion has charge $e/2$. The total electric charge is conserved by looking at $\dot Q_{EM}(t,\infty)$:
\bea
\label{Eq: EM Charge Enclosed}
\dot Q_{EM}(t,\infty)=\sum_i\frac{\dot\phi_i(\infty,t)-\dot\phi_i(0,t)}{4\pi}
\eea
From the Lagrangian in Eq.~\ref{Eq: Bosonized Lagrangian} one can deduce that the only solutions with finite energy are those with $\phi_i(r=\infty)=2\pi n_i$ where $n_i$ is an integer and $\frac{1}{2}\sum_i\phi_i=\theta$. Thus, Eq.~\ref{Eq: EM Charge Enclosed} reduces to $\dot Q_{EM}(t,\infty)=0$ reflecting that electromagnetic charge is conserved in both the $SO(N_f)$ and $Sp(N_f)$ theories. Similarly, from the boundary conditions in Eq.~\ref{Eq: SpN scalar BC and mass} one can see that the $Sp(N_f)$ theory contains $N_f/2$ conserved $B-L$-type currents, $j_{b,i}^\mu-j_{\ell,i}^\mu$, whereas the $SO(N_f)$ does not contain any such currents.

\subsection{2D Fermion Identities}
\label{Appendix: Fermions and Monopoles}
Here we prove the correspondences between terms in the 1+3D theory with those in the 1+1D theory given in Eq.~\ref{Eq: 2D kinetic}-~\ref{Eq: Monopole Term}. Before looking at each term individually, it is first useful to note that from the definition of the 1+1D spinor, $\xi$, and $\bar\gamma^\mu$ in Eq.~\ref{Eq: 2D Spinor and gammas},
\begin{align}
\label{2D current gp identities}
\bar\xi \bar \gamma^0\xi=&\frac{|g-p|^2+|g+p|^2}{2}=g^*g+p^*p\\\nn
\bar\xi \bar \gamma^1\xi=&\frac{|g-p|^2-|g+p|^2}{2}=-(g^*p+p^*g)\\ \nn
\bar\xi \bar \gamma^5\xi=&\frac{(g^*-p^*)(g+p)-(g^*+p^*)(g-p)}{2}=g^*p-p^*g\end{align}
Also, it is easy to see that the spin-isospin singlet state $|s\rangle \equiv |\vec \tau+\vec \sigma = 0\rangle$ obeys the following identities for the spin operator $\vec \sigma$ and the isospin operator $\vec \tau$
\begin{align}
\bra s|\hat{\vec r}\cdot \vec \sigma|s\ket=\bra s|\hat{\vec r}\cdot \vec \tau|s\ket&=0\\\nn
\bra s|(\hat{\vec r}\cdot \vec \sigma)(\hat{\vec r}\cdot \vec \tau)|s\ket&=-1\\
\nn (\hat{\vec r}\cdot \vec \sigma)^2=(\hat{\vec r}\cdot \vec \tau)^2&= \identity
\end{align}
With these identities, it is easy to prove Eq.~\ref{Eq: 2D kinetic}-~\ref{Eq: Monopole Term}. Starting with Eq~\ref{Eq: 2D kinetic} and inserting~\ref{Eq: J=0 Fermions}, we obtain
\bea 
\label{Eq: 2D kinetic proof time}
i\psi^\dag \partial_0\psi=\frac{i}{4\pi r^2}\bra s|(g^*+(\hat{\vec r}\cdot \vec \sigma)p^*)(\dot g+(\hat{\vec r}\cdot \vec \sigma)\dot p)|s\ket=i\frac{g^*\dot g+p^*\dot p}{4\pi r^2}=\frac{i}{4\pi r^2}\bar\xi\bar\gamma^0\partial_0\xi
\eea
and 
\begin{eqnarray}
-i\psi^\dag \sigma_i\partial_i\psi&=&\frac{-i}{\sqrt{4\pi r^2}}\bra s|(g^*+p^*(\hat{\vec r}\cdot \vec \sigma))\sigma_i\partial_i\(\frac{g(r,t)+(\hat{\vec r}\cdot \vec \sigma)p(r,t)}{\sqrt{4\pi r^2} }\)|s\ket\\\nn 
&=&\frac{-i}{4\pi r^2}\langle s|(g^*+p^*(\hat{\vec r}\cdot \vec \sigma))\((\hat{\vec r}\cdot \vec \sigma)g'(r,t)+p'(r,t)-\frac{1}{r}\((\hat{\vec r}\cdot \vec \sigma)g(r,t)-p(r,t)\)\)|s\rangle\\
\nn&=&\frac{-i}{4\pi r^2}\((g^*p'+p^*g')+\frac{1}{r}(g^*p-p^*g)\)=\frac{i}{4\pi r^2}\(\bar \xi \bar \gamma^1\partial_1\xi -\frac{1}{r}\bar \xi \bar\gamma^5\xi\) ,
\end{eqnarray}
which proves Eq.~\ref{Eq: 2D kinetic}. Eqs.~\ref{Eq: Current 2D reduction} and~\ref{Eq: EM Current 2D reduction} can easily be proven from the following identities
\bea 
\label{Eq: 2D kinetic proof time}
\psi^\dag\psi=\frac{1}{4\pi r^2}\bra s|(g^*+(\hat{\vec r}\cdot \vec \sigma)p^*)(g+(\hat{\vec r}\cdot \vec \sigma)p)|s\ket=\frac{g^* g+p^*p}{4\pi r^2}=\frac{1}{4\pi r^2}\bar\xi\bar\gamma^0\xi
\eea
\bea
-\psi^\dag(\hat{\vec r}\cdot \vec \sigma)\psi=-\frac{1}{4\pi r^2}\bra s|(g^*+(\hat{\vec r}\cdot \vec \sigma)p^*)(\hat{\vec r}\cdot \vec \sigma)(g+(\hat{\vec r}\cdot \vec \sigma)p)|s\ket=-\frac{g^* p+p^*g}{4\pi r^2}=\frac{1}{4\pi r^2}\bar\xi\bar\gamma^1\xi
\eea
\bea
\psi^\dag(\hat{\vec r}\cdot \vec \tau)\psi=\frac{1}{4\pi r^2}\bra s|(g^*+(\hat{\vec r}\cdot \vec \sigma)p^*)(\hat{\vec r}\cdot \vec \tau)(g+(\hat{\vec r}\cdot \vec \sigma)p)|s\ket=-\frac{g^* p+p^*g}{4\pi r^2}=\frac{1}{4\pi r^2}\bar\xi\bar\gamma^1\xi
\eea
\bea
-\psi^\dag(\hat{\vec r}\cdot \vec \sigma)(\hat{\vec r}\cdot \vec \tau)\psi=\psi^\dag\psi=\frac{1}{4\pi r^2}\bar\xi\bar\gamma^0\xi
\eea
Finally, we prove Eq.~\ref{Eq: Monopole Term}. For ease of computation, we adopt a coordinate system where $\vec r$ points in the $z$-direction from which one can see 
\bea
\epsilon_{iab}\sigma_i\tau_a\hat r_b=\sigma_x \tau_y-\sigma_y\tau_x .
\eea
Using some simple properties of 2-spin systems, it is easy to show that 
\begin{align}
\sigma_x\tau_y| s\rangle=&-\sigma_y\tau_x| s\rangle \\ \nn \bra s|(g^*+(\hat{\vec r}\cdot \vec \sigma)p^*)(\sigma_x\tau_y)(g+(\hat{\vec r}\cdot \vec \sigma)p)| s\ket=&i\frac{(g^*-p^*)(g+p)-(g^*+p^*)(g-p)}{2}=i\bar\xi\bar\gamma^5\xi .
\end{align}
Then, it is easy to see that
\bea-e\psi^\dag \sigma_i\frac{\tau^a}{2}\psi A_i^a=\frac{A(r)}{2r}\psi^\dag\epsilon_{iab}\sigma_i\tau_a\hat r_b\psi=i\frac{A(r)}{(4\pi r^2)r}\bar \xi\bar\gamma^5\xi .
\eea

\subsection{Boundary Conditions} \label{Appendix: Boundary}
Here we derive the boundary condition $p(r=0)=0$ from the Lagrangian in Eq.~\ref{Eq: 2D Fermion Lagrangian 1}. There we can find the Dirac equation near $r=0$ to be 
\bea
\label{Eq: Dirac Eq monopole}
0=\slashed\partial\xi -\frac{\gamma^5}{r}\xi .
\eea
We have neglected the electromagnetic term since we can use the remaining gauge freedom to set $\dot \lambda(r=0)=0$ and it stays zero near the origin as it take a huge amount of energy to excite the electric field so close to the origin.  The mass term is also negligible since we are looking at scales of order $r_M^{-1}\gg m$. We have also set $A(r)=0$ as we are inside the monopole's core. We can consider time-independent solutions since we are interested in fermions with energies much smaller than $r_{M}^{-1}$. In terms of $g$ and $p$, these equations can be greatly simplified to 
\bea
g'=\frac{g}{r}\qquad p'=-\frac{p}{r} .
\eea
These can easily be seen to have power-law solutions $g\propto r$ and $p\propto r^{-1}$. So we see that to have well-behaved finite solutions near the core we must enforce $p(r=0)=0$.

\section{Bosonization} \label{App: boson}
Here we give a detailed derivation of the bosonization process. We start by proving bosonization for a single fermion on the half-plane by assuming that a commutator of the corresponding scalar fields takes a specific form for small space-like separations. Then we will show that if the fermions in our theory obey the boundary conditions given in Eq.~\ref{Eq: xi BC}-~\ref{Eq: current BC}, the scalar commutator takes the assumed form for both $SO(N_f)$ and $Sp(N_f)$ type boundary conditions. 

\subsection{Bosonization}
\label{Appendix: Bosonization}

Let $\xi$ be a fermion field defined on a half-plane with some boundary conditions at $r=0$ with a Lagrangian of the form
\bea
\label{Eq: Fermion Action}
L=\int_0^\infty dr \;i\bar \xi\slashed\partial \xi-m_i\bar\xi\xi +e \dot\lambda\bar\xi\bar\gamma^0\xi.
\eea
We will be treating the gauge field perturbatively and thus drop the $\dot \lambda$ piece for the rest of this section.
 We will take our basis of $\bar\gamma^\mu$ that is given in Eq.~\ref{Eq: J=0 Fermions}.
In the 1960's, works by Sugawara~\cite{PhysRev.170.1659}, Sommerfield~\cite{PhysRev.176.2019} and others~\cite{PhysRev.165.1883} showed that the stress-energy tensor for a theory of free massless fermions can be written in terms of its current $j^\mu$
\bea
\label{Eq: Sugawara Sress Energy}
T^{\mu\nu}=\frac{\pi}{2}\( \{j^\mu ,j^\nu\}-\eta^{\mu\nu}j^\alpha j_\alpha\)
 \where j^\mu(r)\equiv \lim_{r
\rar r'}\frac{1}{2}\( \bar \xi(r)\bar \gamma^\mu\xi(r')+\bar \xi(r')\bar \gamma^\mu\xi(r)\)\eea
This is known as the Sugawara-Sommerfield construction of the stress-energy tensor. We can use this to write the stress-energy tensor of our theory as
\bea
\label{Eq: Current Hamilton}
T^{\mu\nu}=\frac{\pi}{2}\(\{j^\mu,j^\nu\}-\eta^{\mu\nu}j^\alpha j_{\alpha}\)+\eta^{\mu\nu} m\bar\xi\xi .
\eea
This stress-energy tensor can be used in conjunction with the commutation identities,
\bea
\label{Eq: Current-fermion Commutations}
[j^\mu(x),\xi(y)]=-\bar\gamma^0\bar\gamma^\mu\xi(x)\delta(x-y)\quad [\bar\xi(x)\xi(x),\xi(y)]=-\bar\gamma^0\xi(x)\delta(x-y),
\eea
which can be derived from the canonical equal time anticommutation identities for $\xi$, to derive the Heisenberg equations of motion for $\xi$

\bea
\label{Eq: Fermion Eq of Mo time}
-i\partial_0\xi_i(r)=\[H,\xi(r)\]
\eea
\bea
\label{Eq: Fermion Eq of Mo space}
-i\partial_1\xi_i(r)=\int_0^\infty dr'\[T^0_{\;\;1}(r'),\xi(r)\]=\frac{\pi}{2}\{j_i^1(r)+\bar\gamma^5j_i^0(r),\xi_i(r)\}
\eea
From this point, bosonization is proved in the following way. We start with an ansatz for our fermion field written in terms of a scalar field $\phi$: 
\bea
\label{Eq: xi ansatz}
\tilde \xi(r,t)\equiv Z^{1/2}(r)\mat :e^{-i\sqrt{\pi}\Phi_1(r,t)}: \\ e^{i\alpha}:e^{-i\sqrt{\pi}\Phi_{-1}(r,t)}:\rix \where \Phi_\lambda(r,t)=-\lambda\phi(r,t)+\int_0^r dx\dot \phi(x,t) ,
\eea
$\lambda=\pm 1$, and the colons indicate the normal ordering of the scalar field operators. It's worth noting that $\alpha$ is an arbitrary phase that is eventually determined by the boundary conditions on $\xi$. We define $\Phi_\lambda^+$ and $\Phi_\lambda^-$ to be the portions of $\Phi_\lambda$ that contain the creation and annihilation operator and likewise for $\phi^\pm$. We assume that $C_{\lambda\lambda'}(r,r')$ defined as the limit
\bea
C_{\lambda\lambda'}(r,r')\equiv\lim_{t\rar t'}[\Phi^{-}_\lambda(r,t),\Phi^{+}_{\lambda'}(r',t')]
\eea 
takes the following form
\bea
\label{Eq: Commutator form}
\lim_{r \rar r'} C_{\lambda\lambda'}(r,r')=W(r,r')+\frac{1}{\pi}\mat -\ln(\epsilon-i(r-r'))&& \ln\(\frac{\mu e^\gamma}{2}\)-F(r)-\frac{ i\pi}{2}\\
\ln\(\frac{\mu e^\gamma}{2}\)-F(r)+\frac{ i\pi}{2}&& -\ln(\epsilon+i(r-r'))\rix 
\eea
$\lambda$ labels the columns and $\lambda'$ labels the rows from $+1$ to $-1$, and $\mu$ is the scale at which normal ordering is performed. $F$ and $W$ are, at this point, arbitrary functions with the only assumption being that $W$ is symmetric in its arguments $W(r,r')=W(r',r)$.  $\epsilon$ comes from the $i \epsilon$ procedure and the precise form of $C_{\lambda\lambda'}(r,r')$ will depend on the boundary conditions imposed on the scalars.
Later in this appendix, we prove that $C_{\lambda\lambda'}$ takes this form in both the $SO(N_f)$ and $Sp(N_f)$ theories. We show that if we take $Z(r)=e^{-\pi W(r,r)}/2\pi$ the following claims are true: 
\begin{enumerate}
\item $\xi$ satisfies the current correspondence 
\bea
\label{Eq: Current Correspondence Apdx}
j^\mu=\frac{\epsilon^{\mu\nu}}{\sqrt{\pi}}\partial_\nu\phi .
\eea
\item $ \xi$ satisfies the spatial Heisenberg equations in Eq.~\ref{Eq: Fermion Eq of Mo space}.  Given the Sugawara-Sommerfield construction, the equations of motions are guaranteed to match once the currents are matched, but it is simple enough of a consistency check to perform.
\item $ \xi$ satisfies the canonical anticommutation relations at equal time 
\bea
\label{Eq: Canoncial anticommutation}
\{\xi_\lambda(r),\xi_\lambda(r')\}=0\qquad \{\xi_\lambda(r),\xi^\dag_\lambda(r')\}=\delta_{\lambda\lambda'}\delta (r-r') .
\eea
\item The operator $\bar\xi(r)\xi(r)$ is equal to 
\bea
\label{Eq: General Mass term}
\bar\xi(r)\xi(r)= - \frac{\mu e^\gamma}{2\pi}e^{-F(r)}:\cos(2\sqrt{\pi}\phi-\alpha):
\eea
\end{enumerate}
With these claims proven, we will have shown that the theory for $\xi$ and the theory for $\phi$ are equivalent provided $\phi$ obeys the Hamiltonian
\bea
\label{Eq: Scalar Lagrangian}
H=\int_0^\infty dr \frac{1}{2}(\dot\phi^2+{\phi'}^2)+m\frac{\mu e^\gamma}{4 \pi}e^{-F(r)}(1-:\cos(2\sqrt{\pi}\phi_i+\alpha_i):)-g\dot\lambda\phi'
\eea

It should be noted that, once we have bosonized the theory, the function $W(r,r')$ is unphysical because it cancels in any operator built out of fermion bilinears when translating to normal ordered scalar operators 
 \bea 
 \xi^\dag_\lambda(r)\xi_{\lambda'}(r)\propto Z(r)e^{\pi [\Phi^-_\lambda(r),\Phi^+_{\lambda'}(r)]}\propto e^{-\pi W}e^{\pi W}=1 .
 \eea
We now prove the above four claims thus proving bosonization.
\subsection{Proving the 4 Claims}
\textbf{Claim 1: Current Correspondence}\\
Here we prove the current correspondence in Eq.~\ref{Eq: Current Correspondence Apdx}. We start by noting from the definition in Eq.~\ref{Eq: Sugawara Sress Energy}
\bea
\label{Eq: j0 expanded}
j^0(r)=\frac{1}{2}\lim_{r\rar r'}(\xi^\dag_{1}(r)\xi_{1}(r')+\xi^\dag_{1}(r')\xi_{1}(r)+\xi^\dag_{-1}(r)\xi_{-1}(r')+\xi^\dag_{-1}(r')\xi_{-1}(r))
\eea
\bea
\label{Eq: j1 expanded}
j^1(r)=\frac{1}{2}\lim_{r\rar r'}(\xi^\dag_{1}(r)\xi_{1}(r')+\xi^\dag_{1}(r')\xi_{1}(r)-\xi^\dag_{-1}(r)\xi_{-1}(r')-\xi^\dag_{-1}(r')\xi_{-1}(r) )
\eea
So we will be interested in the combinations 
 $\xi_\lambda^\dag(r)\xi_\lambda(r')$
in the $r\rar r'$ limit. We can use Eq.~\ref{Eq: xi ansatz} to write in the $ r \rar r'$ limit:
\begin{align}
&\xi^\dag_{\lambda}(r)\xi_{\lambda}(r')=Z(r) :e^{-i\sqrt{\pi}(\Phi_\lambda(r')-\Phi_\lambda(r))}:e^{\pi\[\Phi^-_\lambda(r),\Phi^+_\lambda(r')\]}\\\nn
=&:e^{-i\sqrt{\pi}(\Phi_\lambda(r')-\Phi_\lambda(r))}:\frac{e^{-\ln(\epsilon-i\lambda (r-r'))}}{2\pi}=\frac{i\lambda}{2\pi(r-r')}(1+i\sqrt{\pi}(r-r')(\partial_0\phi-\lambda\partial_1\phi))
\end{align}
Now, taking the $r\rar r'$ symmetric limit gives
$$\frac{\xi^\dag_{\lambda}(r)\xi_{\lambda}(r')+\xi^\dag_{\lambda}(r')\xi_{\lambda}(r)}{2}=\frac{\partial_1\phi-\lambda\partial_0\phi}{2\sqrt{\pi}}.$$
At this point we can set $r=r'$. Plugging this into Eqs.~\ref{Eq: j0 expanded} and~\ref{Eq: j1 expanded}, we can see we get the current correspondence in Eq.~\ref{Eq: Current Correspondence Apdx}.

\textbf{Claim 2: $\tilde \xi$ satisfies Eq.~\ref{Eq: Fermion Eq of Mo space}}\\
Now let us prove that $\tilde \xi$ satisfies Eq. ~\ref{Eq: Fermion Eq of Mo space}. From Eq.~\ref{Eq: xi ansatz}: 
\begin{align}-i\partial_r \xi_\lambda=&-\sqrt{\pi Z(r)} e^{-i\sqrt{\pi}\Phi^+_\lambda}(\partial_r\Phi_\lambda)e^{-i\sqrt{\pi}\Phi^-_\lambda}-\frac{i}{2}\partial_r\ln(Z(r))\xi_\lambda\\
\nn =&-\frac{\sqrt{\pi Z(r)}}{{2}}\{\partial_r\Phi_\lambda,:e^{-i\sqrt{\pi}\Phi_\lambda}:\}-\frac{i}{2}\partial_r\ln(Z(r))\xi_\lambda\\ \nn&-\frac{\sqrt{\pi Z(r)}}{{2}}\(\[e^{-i\sqrt{\pi}\Phi^+_\lambda},\partial_r\Phi_\lambda\]e^{-i\sqrt{\pi}\Phi^-_\lambda}+e^{-i\sqrt{\pi}\Phi^+_\lambda}\[\partial_r\Phi_\lambda,e^{-i\sqrt{\pi}\Phi^-_\lambda}\]\) .
\end{align}
There is an extra factor of $e^{i \alpha}$ that is reabsorbed back into $\xi_\lambda$ at Eq.~\ref{Eq: intermediate} when $\lambda = -1$.
It is easy to show that for any two operators $A,B$ whose commutator is a complex number that $[A,e^B]=[A,B]e^B$ and so, 
\bea \label{Eq: intermediate}
-i\partial_r \xi_\lambda=-\frac{\sqrt{\pi}}{{2}}\{\partial_r\Phi_\lambda,\xi_\lambda\}-\frac{i}{2}\partial_r\ln(Z(r))\xi_\lambda-i\frac{\pi}{2}\(\[\partial_r\Phi^-_\lambda,\Phi^+_\lambda\]+\[\Phi^-_\lambda,\partial_r\Phi^+_\lambda\]\)\xi_\lambda
\eea
This last term can be rewritten as
\bea 
\[\partial_r\Phi^-_\lambda,\Phi^+_\lambda\]+\[\Phi^-_\lambda,\partial_r\Phi^+_\lambda\]=\lim_{r\rar r'}(\partial_r+\partial_{r'})\[\Phi^-_\lambda(r),\Phi^+_\lambda(r')\] .
\eea
Using Eq.~\ref{Eq: Commutator form}, this becomes
\begin{align}
\[\partial_r\Phi^-_\lambda,\Phi^+_\lambda\]+\[\Phi^-_\lambda,\partial_r\Phi^+_\lambda\]&\\ \nn
=\lim_{r \rar r'}(\partial_r+\partial_{r'}) ( W(r,r')&-\frac{1}{\pi}\ln(\epsilon-i\lambda(r-r'))) .
\end{align}
The last term is zero since it depends on the combination $r-r'$ and so we find 
$[\partial_r\Phi^-_\lambda,\Phi^+_\lambda]+[\Phi^-_\lambda,\partial_r\Phi^+_\lambda]=2W'(r,r)$
and we arrive at 
\bea 
-i\partial_r \xi_\lambda=-\frac{\sqrt{\pi}}{2}\{\partial_r\Phi_\lambda,\xi_\lambda\}-\frac{i}{2}\partial_r(\ln(Z(r))+2\pi W'(r,r))\xi_\lambda .
\eea
Since $Z(r)\propto e^{-\pi W(r,r)}$ and $W(r,r')=W(r',r)$, it is easy to $\partial_r\ln(Z(r))=-2\pi W'(r,r)$
so this second term is zero. The first term can be simplified using Eq.~\ref{Eq: Current Correspondence Apdx} and we are left with 
$$-i\partial_r \xi_\lambda=\frac{\pi}{2}\{j^1+\lambda j^0,\xi_\lambda\}$$
which for $\lambda=\pm 1$ gives both components of Eq.~\ref{Eq: Fermion Eq of Mo space}

\textbf{Claim 3: Anticommutation Relations}\\
In order to simplify the proof of the commutation relations, let us first prove the following two identities:
\bea
\label{Eq: Anticom ID 1}
e^{-\pi[\Phi^-_\lambda(r),\Phi^+_{\lambda'}(r')]}+e^{-\pi[\Phi^-_{\lambda'}(r'),\Phi^+_\lambda(r)]}=0
\eea
\bea
\label{Eq: Anticom ID 2}
e^{\pi[\Phi^-_\lambda(r),\Phi^+_{\lambda'}(r')]}+e^{\pi[\Phi^-_{\lambda'}(r'),\Phi^+_\lambda(r)]}=Z^{-1}(r)\delta_{\lambda\lambda'}\delta(r-r')
\eea
Starting with Eq.~\ref{Eq: Anticom ID 1}, we can note from Eq.~\ref{Eq: Commutator form}
\bea
\pi[\Phi^-_\lambda(r),\Phi^+_{-\lambda}(r')]=\pi W(r,r')-F(r)+\ln\(\frac{\mu e^\gamma}{2}\)-\frac{i\lambda \pi}{2}
\eea
so that \bea
e^{\pm\pi[\Phi^-_\lambda(r),\Phi^+_{-\lambda}(r')]}+e^{\pm\pi[\Phi^-_{-\lambda}(r'),\Phi^+_{\lambda}(r)]}=e^{\pm(\pi W(r,r')-F(r))}\((\pm i)^\lambda+(\mp i)^\lambda\)=0
\eea
which proves Eqs.~\ref{Eq: Anticom ID 1} and~\ref{Eq: Anticom ID 2} for $\lambda=-\lambda'$. To prove them for $\lambda=\lambda'$, we again start with Eq.~\ref{Eq: Commutator form} to see
\bea
\pi[\Phi^-_\lambda(r),\Phi^+_{\lambda}(r')]=\pi W(r,r')-\ln(\epsilon-i\lambda(r-r')) ,
\eea
from which it is easy to see
\begin{align}
e^{\pi[\Phi^-_\lambda(r),\Phi^+_{\lambda}(r')]}+e^{\pi[\Phi^-_{\lambda}(r'),\Phi^+_{-\lambda}(r)]}=&2\pi e^{\pi W(r,r')}\delta(r-r')\\ \nn
e^{-\pi[\Phi^-_\lambda(r),\Phi^+_{\lambda}(r')]}+e^{-\pi[\Phi^-_{\lambda}(r'),\Phi^+_{-\lambda}(r)]}=&2\epsilon e^{- \pi W(r,r')}=0
\end{align}
which now proves Eq.~\ref{Eq: Anticom ID 1} and~\ref{Eq: Anticom ID 2} for all $\lambda,\lambda'$.

From Eq.~\ref{Eq: Anticom ID 1} and~\ref{Eq: Anticom ID 2}, it is very easy to compute the anticommutation relations. Note, we can suppress all factors of $e^{i\alpha}$ here since the commutators in which they don't cancel with their conjugate $e^{-i\alpha}$ are zero.  Similarly, we will use the subsitution $\sqrt{Z(r) Z(r')} \rightarrow Z(r)$ as the only non-zero term has $\delta(r-r')$. With this in mind
\begin{align}
\{\xi_\lambda(r),\xi_{\lambda'}(r')\}=&Z(r)\{:e^{-i\sqrt{\pi}\Phi_\lambda(r)}:,:e^{-i\sqrt{\pi}\Phi_{\lambda'}(r')}:\}\\ \nn
=&Z(r):e^{-i\sqrt{\pi}(\Phi_\lambda(r)+\Phi_{\lambda'}(r'))}:\(e^{-\pi[\Phi^-_\lambda(r),\Phi^+_{\lambda'}(r')]}+e^{-\pi[\Phi^-_{\lambda'}(r'),\Phi^+_\lambda(r)]}\)=0
\end{align}
and 
\begin{align}
\{\xi_\lambda(r),\xi^\dag_{\lambda'}(r')\}=&Z(r)\{:e^{-i\sqrt{\pi}\Phi_\lambda(r)}:,:e^{i\sqrt{\pi}\Phi_{\lambda'}(r')}:\}\\
\nn=&Z(r):e^{-i\sqrt{\pi}(\Phi_\lambda(r)-\Phi_{\lambda'}(r'))}:\(e^{\pi[\Phi^-_\lambda(r),\Phi^+_{\lambda'}(r')]}+e^{\pi[\Phi^-_{\lambda'}(r'),\Phi^+_\lambda(r)]}\)\\ \nn
=&Z(r):e^{-i\sqrt{\pi}(\Phi_\lambda(r)-\Phi_{\lambda'}(r'))}:\frac{\delta_{\lambda\lambda'}\delta(r-r')}{Z(r)}=\delta_{\lambda\lambda'}\delta(r-r')
\end{align}
which proves the commutation relations. 

\textbf{Claim 4: Mass Term Correspondence}\\
Next let us prove the correspondence for the mass term in Eq.~\ref{Eq: General Mass term}.  At this point, we simply compute
\begin{align}
\bar\xi(r)\xi(r)=&Z(r)\(ie^{-i\alpha}:e^{i\sqrt{\pi}\Phi_{-1}(r)}::e^{-i\sqrt{\pi}\Phi_{1}(r)}:-i e^{i \alpha} : e^{i\sqrt{\pi}\Phi_{1}(r)}::e^{-i\sqrt{\pi}\Phi_{-1}(r)}:\)\\ \nn
=&Z(r)\(ie^{-i\alpha}:e^{2i\sqrt{\pi}\phi}:e^{\pi\[\Phi_{-1}^-(r),\Phi_{1}^+(r)\]}-ie^{i\alpha}:e^{-2i\sqrt{\pi}\phi}:e^{\pi\[\Phi_1^-(r),\Phi_{-1}^+(r)\]}\)
\end{align}
where we have used $\Phi_{-1}-\Phi_1=2\phi$. These commutators have a well-defined limit at $r=r'$.
\bea\pi\[\Phi_\lambda^-(r),\Phi_{-\lambda}^+(r)\]=\pi W(r,r)-F(r)+\ln\(\frac{\mu e^\gamma}{2}\)-\frac{i\lambda\pi}{2}\eea
So, 
\begin{eqnarray}
\bar\xi(r)\xi(r)&=&-Z(r)\frac{\mu e^{\pi W(r,r)+\gamma-F(r)}}{2}\(:e^{i(2\sqrt{\pi}\phi-\alpha)}:+:e^{-i(2\sqrt{\pi}\phi-\alpha)}:\)  \nn \\
&=&-\frac{\mu e^{\gamma-F(r)}}{2\pi}:\cos\(2\sqrt{\pi}\phi-\alpha\)
\end{eqnarray}
This proves Eq.~\ref{Eq: General Mass term}. 

\subsection{Commutator Computation}
The only remaining task is to show that $C_{\lambda\lambda'}$ takes the supposed form in Eq.~\ref{Eq: Commutator form} for both $SO(N_f)$ and $SU(N_f)$ theories. From Eq.~\ref{Eq: xi ansatz}, we can see that $C_{\lambda\lambda'}$ takes the form
\begin{align}
\label{Eq: Commutator start}
 C_{\lambda\lambda'}(r,r')&=\lim_{t\rar t',r\rar r'}\bigg(\int_0^rdx \int_0^{r'} dx'\partial_t\partial_{t'}D(x,t;,x',t')\\ \nn&-\lambda\int_0^{r'}dx'\partial_{t'}D(r,t;x',t')-\lambda'\int_0^{r}dx\partial_{t}D(x,t;r',t')+\lambda\lambda'D(r,t;r',t')\bigg) ,
 \end{align}
where $D(r,t;r',t')\equiv[\phi^-(r,t),\phi^+(r',t')]$. Before diving into a computation of $C_{\lambda\lambda'}$, we first give a brief argument that we can compute $C_{\lambda\lambda'}$ by replacing the full propagator $D$ with the equivalent propagator for a free massless scalar $D_0$. Let us argue this term by term in Eq.~\ref{Eq: Commutator start}. As we will see, the first term will be entirely absorbed into $W(r,r')$ and so will not concern us. If we assume that our theory is perturbative, the full propagator $D$ can be approximated by the propagator to arbitrarily high accuracy by the free field propagator $D_0$ in the $t\rar t'$ and $r\rar r'$ limit and so we can make the replacement in the last term. We can also replace $D$ with $D_0$ in the middle two terms provided that $\lim_{t\rar t'}\partial_tD(r,t;,r',t')=0$ if $r\neq r'$ so that this integral receives no contribution at finite $x-r'$ (or $r-x'$). To see this it true, note that from time translation and time reversal symmetry, $D(r,t,r',t')$ can only depend on $(t-t')^2$ and thus its time derivative must be 0 or ill-defined when $t=t'$. Because $D$ is the Green's function of our full theory, it must be continuous and smooth when $t\neq t'$ and $r\neq r'$ so this time derivative can only be ill-defined when $r=r'$ and $t=t'$. So we find that $\lim_{t\rar t'}\partial_tD(r,t;,r',t')=0$ when $r\neq r'$ and thus we can replace $D$ with the free scalar propagator $D_0$ in the middle two terms as well. With these replacements, $C_{\lambda\lambda'}$ takes the form 
\begin{align}
\label{Eq: Commutator next}
 \lim_{r\rar r'}C_{\lambda\lambda'}(r,r')=&\lim_{t\rar t',r\rar r'}\bigg(\lambda\lambda'D_0(r,t;r',t')-\lambda\int_0^{r'}dx'\partial_{t'}D_0(r,t;x',t')\\ \nn&-\lambda'\int_0^{r}dx\partial_{t}D_0(x,t;r',t')+\tilde W(r,r')
 \end{align}
where 
\bea
\tilde W(r,r')=\int_0^rdx \int_0^{r'} dx'\partial_t\partial_{t'}D(x,t;,x',t') .
\eea
Now that we have rewritten $C_{\lambda\lambda'}$ in terms of the free scalar propagator, $D_0$, we have computational control and we can compute this for the $SO(N_f)$ and $Sp(N_f)$ theories. Because these two theories have different boundary conditions, $D_0$ will be different for the two theories. We will compute $C_{\lambda\lambda'}$ in both theories. Before proceeding it is worth commenting on a technical aspect of free, massless scalar theory in 1+1D. These theories contain IR divergences which fortunately can be regulated entirely through normal ordering. Normal ordering necessitates the introduction of a scale $\mu$ with respect to which normal ordering is performed~\cite{Coleman:1974bu}. Thus a theory of free, massless scalar fields in 1+1D must include the specification of a scale $\mu$ if it is to be well-defined. This scale $\mu$ appears as a "mass" for the scalar field but it should be noted that it is not meant to represent the mass of any particle. 

\textbf{$SO(N_f)$ Theories}\\
The bosonized theory for $SO(N_f)$ theories consists of $N_f$ scalar fields, $\phi_i$, the boundary conditions for which can be found by applying the boundary conditions in Eq.~\ref{Eq: current BC} to the current correspondence in Eq.~\ref{Eq: Current Correspondence Apdx}
\bea
\partial_r\phi_i(r=0)=0 .
\eea
One can easily see that the free, massless scalar fields with this boundary condition when quantized take the form 
\bea
\label{Eq: Quantized scalar field}
\phi(r,t)=\int_{0}^\infty \frac{dk}{2\pi\sqrt{2 \omega}} 2 \cos(kr)e^{-i\omega t}a_k +h.c.
\eea
where $\omega=\sqrt{k^2+\mu^2}$ and $a_k(a^\dag_k)$ are the creation and annihilation operators. From this, we can easily compute $D_0$
\begin{align}
D_0(r,t;r't')&\equiv[\phi^-(r,t),\phi^+(r',t')]=\int_{-\infty}^\infty\frac{dk}{2\pi\omega}e^{-i\omega(t-t')}\cos(kr)\cos(kr')\\
\nn &=\int_{-\infty}^\infty\frac{dk}{4\pi\omega}\left( e^{-i(\omega(t-t')-k(r-r'))}+e^{-i(\omega(t-t')-k(r+r'))}\right)\\ \nn&=G_0(r-r',t-t')+G_0(r+r',t-t')
\end{align}
where
\bea
\label{Eq: G0 def}
G_0(\Delta r,\Delta t)\equiv\int_{-\infty}^\infty\frac{dk}{4\pi\omega} e^{-i(\omega\Delta t-k\Delta r)}=\frac{1}{2\pi}K_0(\mu\sqrt{\Delta r^2-(\Delta t-i\epsilon)^2}) .
\eea
It is understood that we take the limit of $\epsilon\rar 0$ at the end of any computation. 
Now, we can plug this into Eq.~\ref{Eq: Commutator start} and compute. Firstly it is not difficult to show that 
\bea
\lim_{t\rar t'}\partial_tG_0(r\pm r',t-t')=-\frac{i\epsilon}{2\pi(\epsilon^2+(r\pm r')^2)}\(\mu\sqrt{(r\pm r')^2+\epsilon^2}K_1(\mu\sqrt{(r\pm r')^2+\epsilon^2})\)
\eea
Clearly this will vanish in the $\epsilon\rar 0$ limit unless $r\pm r'=0$ so we can expand around this point to find 
\bea
\lim_{\epsilon\rar 0,t\rar t'}\partial_tG_0(r\pm r',t-t')=-\frac{i\epsilon}{2\pi(\epsilon^2+(r\pm r')^2)} ,
\eea
which can be quickly integrated to give
\bea
\lim_{\epsilon\rar 0,t\rar t'}\int_0^r dx\partial_tG_0(x\pm r',t-t')=-\frac{i}{2\pi}\(\arctan\(\frac{r\pm r'}{\epsilon}\)-\arctan\(\frac{\pm r'}{\epsilon}\)\) .
\eea
This can be greatly simplified by noting that since $r,r'>0$, in the $\epsilon\rar 0$ many of these $\arctan$ factors reduce to $\pm\pi/2$ depending on the sign of the argument. When the dust settles we find 
\bea
\lim_{\epsilon\rar 0,t\rar t'}\int_0^r dx\partial_tD_0(x,t;r',t')=-\frac{i}{2\pi}\(\arctan\(\frac{r- r'}{\epsilon}\)+\frac{\pi}{2}\) .
\eea
A nearly identical computation shows that 
\bea
\lim_{\epsilon\rar 0,t\rar t'}\int_0^{r'} dx'\partial_{t'}D_0(r,t;x',t')=-\frac{i}{2\pi}\(\arctan\(\frac{r- r'}{\epsilon}\)-\frac{\pi}{2}\) .
\eea
This combined with the limits
\begin{align}
\lim_{r-r',t-t',\epsilon\rar 0}G_0(r-r',t-t')&=-\frac{1}{2\pi}\ln\(\frac{\mu e^\gamma}{2}\sqrt{(r-r')^2+\epsilon^2}\)\\ \nn\lim_{r-r',t-t',\epsilon\rar 0}G_0(r+r',t-t')&=\frac{K_0(\mu (r+r'))}{2\pi}
\end{align}
allows us to write
\begin{align}
\label{Eq: C SON}
C_{\lambda\lambda'}(r,t;,r',t')=&-\frac{\lambda\lambda'}{2\pi}\(\ln\(\frac{\mu e^\gamma}{2}\sqrt{(r-r')^2+\epsilon^2}\)-K_0(\mu(r+r'))\)\\
&+\frac{i(\lambda+\lambda')}{2\pi}\arctan\(\frac{r-r'}{\epsilon}\)-\frac{i(\lambda-\lambda')}{4}+\tilde W(r,r) .
\end{align}
Finally, if we make the definition
\bea
W(r,r')\equiv\tilde W(r,r')+\frac{1}{2\pi}\(\ln(\sqrt{(r-r')^2+\epsilon^2})-\ln\(\frac{\mu e^\gamma}{2}\)+K_0(\mu (r+r'))\)
\eea
and expand the $\arctan$ out in terms of logs, one can show that this takes the form 
\begin{align}
C_{\lambda\lambda'}(r,t;,r',t)=&-\frac{(1+\lambda)(1+\lambda')}{4\pi}\ln(\epsilon-i(r-r'))-\frac{(1-\lambda)(1-\lambda')}{4\pi}\ln(\epsilon+i(r-r'))\\\nn &-\frac{1-\lambda\lambda'}{2\pi}\(K_0(2\mu r)-\ln\(\frac{\mu e^\gamma}{2}\)\)-\frac{i(\lambda-\lambda')}{4}+W(r,r) ,
\end{align}
which can be seen to be exactly the form given in Eq.~\ref{Eq: Commutator form} with $F(r)=K_0(2\mu r)$. 

\textbf{$Sp(N_f)$ Theories}\\
The $Sp(N_f)$ theories consist of $N_f/2$ families of 2 scalar fields $\phi_{b,i}$ and $\phi_{\ell,i}$. Since the full theory consists of $N_f/2$ identical families, we can focus on one family and drop the $i$ index. We wish to quantize these free, massless fields. In general, we can write them in terms of creation and annihilation operators $a(a^\dag)$ and $b(b^\dag)$ for fields $\phi_\ell$ and $\phi_b$ respectively, separating out the left and right moving components. 
\begin{align}
\label{Eq: Scalar fields}
&\phi_\ell(r,t)=\int_0^\infty \frac{dk}{2\pi\sqrt{2\omega}}e^{-i\omega t}(a_k e^{ikr}+a_{-k}e^{-ikr})+h.c.\\
&\phi_b(r,t)=\int_0^\infty \frac{dk}{2\pi\sqrt{2\omega}}e^{-i\omega t}(b_k e^{ikr}+b_{-k}e^{-ikr})+h.c.
\end{align}
Note that each scalar field has its own scale $\mu$. On these fields we can impose the boundary conditions.
\bea
\partial_r(\phi_\ell+\phi_b)=0 \quad \text{and} \quad \partial_t(\phi_\ell-\phi_b)=0
\eea 
derived from combining Eq.~\ref{Eq: current BC} and~\ref{Eq: xi ansatz}. These boundary conditions can easily be seen to enforce 
\bea
\label{Eq: BC on Creation/Annihilation Ops}
 a_{-k}=b_{k} \quad b_{-k}=a_{k} ,
\eea
so we can write our two fields as
\begin{align}
\label{Eq: SpN quantized Scalar fields}
&\phi_\ell(r,t)=\int_0^\infty \frac{dk}{2\pi\sqrt{2\omega}}e^{-i\omega t}(a_k e^{ikr}+b_{k}e^{-ikr})+h.c.\\
&\phi_b(r,t)=\int_0^\infty \frac{dk}{2\pi\sqrt{2\omega}}e^{-i\omega t}(b_k e^{ikr}+a_{k}e^{-ikr})+h.c.
\end{align}
From this, we can compute the propagator for each field, which, because the labels $a$ and $b$ are arbitrary, are identical. 
\bea\label{Eq: SpN Prop}
D_0(r,t;r',t')=\int_{-\infty}^\infty\frac{dk}{4\pi\omega}e^{-i(\omega(t-t')-k(r-r'))}= G_0(r-r',t-t') .
\eea
This is the same propagator as for $SO(N_f)$ theories but without the $G_0(r+r',t-t')$ piece. The computation of $C_{\lambda \lambda'}$ is then a simplified version of the computation for the $SO(N_f)$ theories and so we simply state the result
\begin{align}
\label{Eq: C SpN}
C_{\lambda\lambda'}(r,t;,r',t')=&-\frac{\lambda\lambda'}{2\pi}\ln\(\frac{\mu e^\gamma}{2}\sqrt{(r-r')^2+\epsilon^2}\)\\
&+\frac{i(\lambda+\lambda')}{2\pi}\arctan\(\frac{r-r'}{\epsilon}\)-\frac{i(\lambda-\lambda')}{4}+\tilde W(r,r) .
\end{align}
If we define
\bea
W(r,r')\equiv\tilde W(r,r')+\frac{1}{2\pi}\(\ln(\sqrt{(r-r')^2+\epsilon^2})-\ln\(\frac{\mu e^\gamma}{2}\)\)
\eea
and expand the $\arctan$ our in terms of logs, one can show that this takes the form 
\begin{align}
C_{\lambda\lambda'}(r,t;,r',t)=&-\frac{(1+\lambda)(1+\lambda')}{4\pi}\ln(\epsilon-i(r-r'))-\frac{(1-\lambda)(1-\lambda')}{4\pi}\ln(\epsilon+i(r-r'))\\\nn &+\frac{1-\lambda\lambda'}{2\pi}\ln\(\frac{\mu e^\gamma}{2\pi}\)-\frac{i(\lambda-\lambda')}{4}+W(r,r) ,
\end{align}
which can be seen to be exactly the form given in Eq.~\ref{Eq: Commutator form} with $F(r)=0$.


\bibliography{biblio}{}

\providecommand{\href}[2]{#2}\begingroup\raggedright\begin{thebibliography}{10}

\bibitem{tHooft:1974kcl}
G.~'t~Hooft, {\it {Magnetic Monopoles in Unified Gauge Theories}},  {\em Nucl.
  Phys. B} {\bf 79} (1974) 276--284.

\bibitem{Polyakov:1974ek}
A.~M. Polyakov, {\it {Particle Spectrum in Quantum Field Theory}},  {\em JETP
  Lett.} {\bf 20} (1974) 194--195.

\bibitem{Callan:1982ah}
C.~G. Callan, Jr., {\it {Disappearing Dyons}},  {\em Phys. Rev. D} {\bf 25}
  (1982) 2141.

\bibitem{Preskill:1984gd}
J.~Preskill, {\it {MAGNETIC MONOPOLES}},  {\em Ann. Rev. Nucl. Part. Sci.} {\bf
  34} (1984) 461--530.

\bibitem{Olaussen:1983qc}
K.~Olaussen, H.~A. Olsen, P.~Osland, and I.~Overbo, {\it {KAZAMA-YANG MONOPOLE
  - FERMION BOUND STATES. 1. ANALYTIC RESULTS}},  {\em Nucl. Phys. B} {\bf 228}
  (1983) 567--587.

\bibitem{Osland:1984yu}
P.~Osland and T.~T. Wu, {\it {Monopole - Fermion and Dyon - Fermion Bound
  States. 1. General Properties and Numerical Results}},  {\em Nucl. Phys. B}
  {\bf 247} (1984) 421--449.

\bibitem{Osland:1984ys}
P.~Osland and T.~T. Wu, {\it {Monopole - Fermion and Dyon - Fermion Bound
  States. 2. Weakly Bound States for the Lowest Angular Momentum}},  {\em Nucl.
  Phys. B} {\bf 247} (1984) 450--470.

\bibitem{Osland:1985va}
P.~Osland and T.~T. Wu, {\it {Monopole - Fermion and Dyon - Fermion Bound
  States. 6. Weakly Bound States for the Dyon - Fermion System}},  {\em Nucl.
  Phys. B} {\bf 261} (1985) 687--730.

\bibitem{Zhang:1988ab}
J.-z. Zhang and Y.-c. Qi, {\it {WAVE FUNCTIONS OF BOUND STATES OF A FERMION AND
  A DIRAC DYON AND MATRIX ELEMENTS IN AN EXTERNAL ELECTROMAGNETIC FIELD}},
  {\em J. Math. Phys.} {\bf 31} (1990) 1796--1799.

\bibitem{Tang:1982fc}
J.-F. Tang, {\it {FERMION BOUND STATES IN A DYON FIELD}},  {\em Phys. Rev. D}
  {\bf 26} (1982) 510--514.

\bibitem{Sen:1994yi}
A.~Sen, {\it {Dyon - monopole bound states, selfdual harmonic forms on the
  multi - monopole moduli space, and SL(2,Z) invariance in string theory}},
  {\em Phys. Lett. B} {\bf 329} (1994) 217--221,
  [\href{http://arxiv.org/abs/hep-th/9402032}{{\tt hep-th/9402032}}].

\bibitem{Sethi:1995zm}
S.~Sethi, M.~Stern, and E.~Zaslow, {\it {Monopole and Dyon bound states in N=2
  supersymmetric Yang-Mills theories}},  {\em Nucl. Phys. B} {\bf 457} (1995)
  484--512, [\href{http://arxiv.org/abs/hep-th/9508117}{{\tt hep-th/9508117}}].

\bibitem{Gauntlett:1995fu}
J.~P. Gauntlett and J.~A. Harvey, {\it {S duality and the dyon spectrum in N=2
  superYang-Mills theory}},  {\em Nucl. Phys. B} {\bf 463} (1996) 287--314,
  [\href{http://arxiv.org/abs/hep-th/9508156}{{\tt hep-th/9508156}}].

\bibitem{Jackiw:1975fn}
R.~Jackiw and C.~Rebbi, {\it {Solitons with Fermion Number 1/2}},  {\em Phys.
  Rev. D} {\bf 13} (1976) 3398--3409.

\bibitem{Callias:1977kg}
C.~Callias, {\it {Index Theorems on Open Spaces}},  {\em Commun. Math. Phys.}
  {\bf 62} (1978) 213--234.

\bibitem{Weinberg:1979ma}
E.~J. Weinberg, {\it {Parameter Counting for Multi-Monopole Solutions}},  {\em
  Phys. Rev. D} {\bf 20} (1979) 936--944.

\bibitem{Callan:1983tm}
C.~g. Callan, Jr., {\it {THE MONOPOLE CATALYSIS S MATRIX}},  in {\em {Workshop
  on Problems in Unification and Supergravity}}, pp.~45--53, 1983.

\bibitem{Witten:1982fp}
E.~Witten, {\it {An SU(2) Anomaly}},  {\em Phys. Lett. B} {\bf 117} (1982)
  324--328.

\bibitem{Callan:1982au}
C.~G. Callan, Jr., {\it {Dyon-Fermion Dynamics}},  {\em Phys. Rev. D} {\bf 26}
  (1982) 2058--2068.

\bibitem{Dawson:1983cm}
S.~Dawson and A.~N. Schellekens, {\it {Monopole - Fermion Interactions: The
  Soliton Picture}},  {\em Phys. Rev. D} {\bf 28} (1983) 3125.

\bibitem{Callan:1983nx}
C.~G. Callan, Jr. and E.~Witten, {\it {Monopole Catalysis of Skyrmion Decay}},
  {\em Nucl. Phys. B} {\bf 239} (1984) 161--176.

\bibitem{Rubakov:1981rg}
V.~A. Rubakov, {\it {Superheavy Magnetic Monopoles and Proton Decay}},  {\em
  JETP Lett.} {\bf 33} (1981) 644--646.

\bibitem{Rubakov:1982fp}
V.~A. Rubakov, {\it {Adler-Bell-Jackiw Anomaly and Fermion Number Breaking in
  the Presence of a Magnetic Monopole}},  {\em Nucl. Phys. B} {\bf 203} (1982)
  311--348.

\bibitem{Rubakov:1983sy}
V.~a. Rubakov and M.~s. Serebryakov, {\it {ANOMALOUS BARYON NUMBER
  NONCONSERVATION IN THE PRESENCE OF SU(5) MONOPOLES}},  {\em Nucl. Phys. B}
  {\bf 218} (1983) 240--268.

\bibitem{Wilczek:1981dr}
F.~Wilczek, {\it {Remarks on Dyons}},  {\em Phys. Rev. Lett.} {\bf 48} (1982)
  1146.

\bibitem{Polchinski:1984uw}
J.~Polchinski, {\it {Monopole Catalysis: The Fermion Rotor System}},  {\em
  Nucl. Phys. B} {\bf 242} (1984) 345--363.

\bibitem{Kitano:2021pwt}
R.~Kitano and R.~Matsudo, {\it {Missing final state puzzle in the
  monopole-fermion scattering}},  {\em Phys. Lett. B} {\bf 832} (2022) 137271,
  [\href{http://arxiv.org/abs/2103.13639}{{\tt arXiv:2103.13639}}].

\bibitem{Csaki:2020inw}
C.~Csaki, S.~Hong, Y.~Shirman, O.~Telem, J.~Terning, and M.~Waterbury, {\it
  {Scattering amplitudes for monopoles: pairwise little group and pairwise
  helicity}},  {\em JHEP} {\bf 08} (2021) 029,
  [\href{http://arxiv.org/abs/2009.14213}{{\tt arXiv:2009.14213}}].

\bibitem{Csaki:2021ozp}
C.~Cs\'aki, Y.~Shirman, O.~Telem, and J.~Terning, {\it {Monopoles Entangle
  Fermions}},  \href{http://arxiv.org/abs/2109.01145}{{\tt arXiv:2109.01145}}.

\bibitem{Brennan:2021ewu}
T.~D. Brennan, {\it {Callan-Rubakov effect and higher charge monopoles}},  {\em
  JHEP} {\bf 02} (2023) 159, [\href{http://arxiv.org/abs/2109.11207}{{\tt
  arXiv:2109.11207}}].

\bibitem{Hamada:2022eiv}
Y.~Hamada, T.~Kitahara, and Y.~Sato, {\it {Monopole-fermion scattering and
  varying Fock space}},  {\em JHEP} {\bf 11} (2022) 116,
  [\href{http://arxiv.org/abs/2208.01052}{{\tt arXiv:2208.01052}}].

\bibitem{vanBeest:2023dbu}
M.~van Beest, P.~Boyle~Smith, D.~Delmastro, Z.~Komargodski, and D.~Tong, {\it
  {Monopoles, Scattering, and Generalized Symmetries}},
  \href{http://arxiv.org/abs/2306.07318}{{\tt arXiv:2306.07318}}.

\bibitem{Brennan:2023tae}
T.~D. Brennan, {\it {A New Solution to the Callan Rubakov Effect}},
  \href{http://arxiv.org/abs/2309.00680}{{\tt arXiv:2309.00680}}.

\bibitem{vanBeest:2023mbs}
M.~van Beest, P.~Boyle~Smith, D.~Delmastro, R.~Mouland, and D.~Tong, {\it
  {Fermion-Monopole Scattering in the Standard Model}},
  \href{http://arxiv.org/abs/2312.17746}{{\tt arXiv:2312.17746}}.

\bibitem{Csaki:2024ajo}
C.~Cs\'aki, R.~Ovadia, O.~Telem, J.~Terning, and S.~Yankielowicz, {\it {Abelian
  Instantons and Monopole Scattering}},
  \href{http://arxiv.org/abs/2406.13738}{{\tt arXiv:2406.13738}}.

\bibitem{Loladze:2024ayk}
V.~Loladze and T.~Okui, {\it {Monopole-Fermion Scattering and the Solution to
  the Semiton/Unitarity Puzzle}},  \href{http://arxiv.org/abs/2408.04577}{{\tt
  arXiv:2408.04577}}.

\bibitem{Vafa:1983tf}
C.~Vafa and E.~Witten, {\it {Restrictions on Symmetry Breaking in Vector-Like
  Gauge Theories}},  {\em Nucl. Phys. B} {\bf 234} (1984) 173--188.

\bibitem{Vafa:1984xg}
C.~Vafa and E.~Witten, {\it {Parity Conservation in QCD}},  {\em Phys. Rev.
  Lett.} {\bf 53} (1984) 535.

\bibitem{Coleman:1974bu}
S.~R. Coleman, {\it {The Quantum Sine-Gordon Equation as the Massive Thirring
  Model}},  {\em Phys. Rev. D} {\bf 11} (1975) 2088.

\bibitem{Mandelstam:1975hb}
S.~Mandelstam, {\it {Soliton Operators for the Quantized Sine-Gordon
  Equation}},  {\em Phys. Rev. D} {\bf 11} (1975) 3026.

\bibitem{PhysRev.170.1659}
H.~Sugawara, {\it A field theory of currents},  {\em Phys. Rev.} {\bf 170}
  (Jun, 1968) 1659--1662.

\bibitem{PhysRev.176.2019}
C.~M. Sommerfield, {\it Currents as dynamical variables},  {\em Phys. Rev.}
  {\bf 176} (Dec, 1968) 2019--2025.

\bibitem{PhysRev.165.1883}
C.~G. Callan, R.~F. Dashen, and D.~H. Sharp, {\it Solvable two-dimensional
  field theory based on currents},  {\em Phys. Rev.} {\bf 165} (Jan, 1968)
  1883--1886.

\end{thebibliography}\endgroup

\bibliographystyle{JHEP}

\end{document}